\documentclass[preprint,12pt]{elsarticle}

\usepackage{amssymb}
\usepackage{amsmath}
\usepackage{array}
\usepackage{makecell}
\usepackage{hyperref}
\usepackage[ruled,vlined]{algorithm2e}

\journal{Computers in Biology and Medicine}

\begin{document}

\begin{frontmatter}



\title{MOIS-SAM2: Exemplar-based Segment Anything Model 2 for multi-lesion interactive segmentation of neurofibromas in whole-body MRI}

\author[uke-iam]{Georgii Kolokolnikov\corref{cor1}}
\ead{g.kolokolnikov@uke.de}
\author[uke-radiology]{Marie-Lena Schmalhofer}
\author[uke-radiology]{Sophie Götz}
\author[uke-radiology]{Lennart Well}
\author[uke-neurology]{Said Farschtschi}
\author[uke-neurology]{Victor-Felix Mautner}
\author[uke-radiology]{Inka Ristow}
\author[uke-iam]{René Werner}

\cortext[cor1]{Corresponding author}

\affiliation[uke-iam]{organization={Institute for Applied Medical Informatics, Institute of Computational Neuroscience, and Center for Biomedical Artificial Intelligence (bAIome), University Medical Center Hamburg-Eppendorf},
            addressline={Martinistr.~52}, 
            city={Hamburg},
            postcode={20246}, 
            country={Germany}}

\affiliation[uke-radiology]{organization={Department of Diagnostic and Interventional Radiology and Nuclear Medicine, University Medical Center Hamburg-Eppendorf},
            addressline={Martinistr.~52}, 
            city={Hamburg},
            postcode={20246}, 
            country={Germany}}

\affiliation[uke-neurology]{organization={Department of Neurology, University Medical Center Hamburg-Eppendorf},
            addressline={Martinistr.~52}, 
            city={Hamburg},
            postcode={20246}, 
            country={Germany}}

%% Abstract
\begin{abstract}
\textbf{Background and Objectives:} Neurofibromatosis type 1 is a genetic disorder characterized by the development of numerous neurofibromas (NFs) throughout the body. Whole-body MRI (WB-MRI) is the clinical standard for detection and longitudinal surveillance of NF tumor growth; however, manual segmentation of these lesions is labor-intensive. Existing interactive segmentation methods fail to combine high lesion-wise precision with scalability to hundreds of lesions. This study proposes a novel interactive segmentation model tailored to this challenge.\\
\textbf{Methods:} We introduce MOIS-SAM2 – a multi-object interactive segmentation model that extends the state-of-the-art, transformer-based, promptable Segment Anything Model 2 (SAM2) with exemplar-based semantic propagation. The model implements user prompts to segment a small set of lesions and propagates this knowledge to similar, unprompted lesions across the entire scan. In this retrospective study, MOIS-SAM2 was trained and evaluated on 119 WB-MRI scans from 84 NF1 patients acquired using T2-weighted fat-suppressed sequences. The dataset was split at the patient level into a training set and four test sets (one in-domain and three reflecting different domain shift scenarios, e.g., MRI field strength variation, low tumor burden, differences in clinical site and scanner vendor). Segmentation performance was assessed using scan-wise Dice Similarity Coefficient (DSC), lesion detection F1 score, and lesion-wise DSC.\\
\textbf{Results:} On the in-domain test set, MOIS-SAM2 achieved a scan-wise DSC of 0.60 against expert manual annotations, outperforming baseline 3D nnU-Net (DSC: 0.54) and SAM2 (DSC: 0.35). Performance of the proposed model was maintained under MRI field strength shift (DSC: 0.53) and scanner vendor variation (DSC: 0.50), and improved in low tumor burden cases (DSC: 0.61). Lesion detection F1 scores ranged from 0.62 to 0.78 across test sets. Preliminary inter-reader variability analysis showed model-to-expert agreement (DSC: 0.62–0.68), comparable to inter-expert agreement (DSC: 0.57–0.69). \\
\textbf{Conclusions:} The proposed MOIS-SAM2 enables efficient and scalable interactive segmentation of NFs in WB-MRI with minimal user input and strong generalization, supporting integration into clinical workflows. The model and code are publicly available on \href{https://github.com/IPMI-ICNS-UKE/MOIS_SAM2_NF}{GitHub}.

\end{abstract}

\begin{graphicalabstract}
\includegraphics[width=1.0\textwidth]{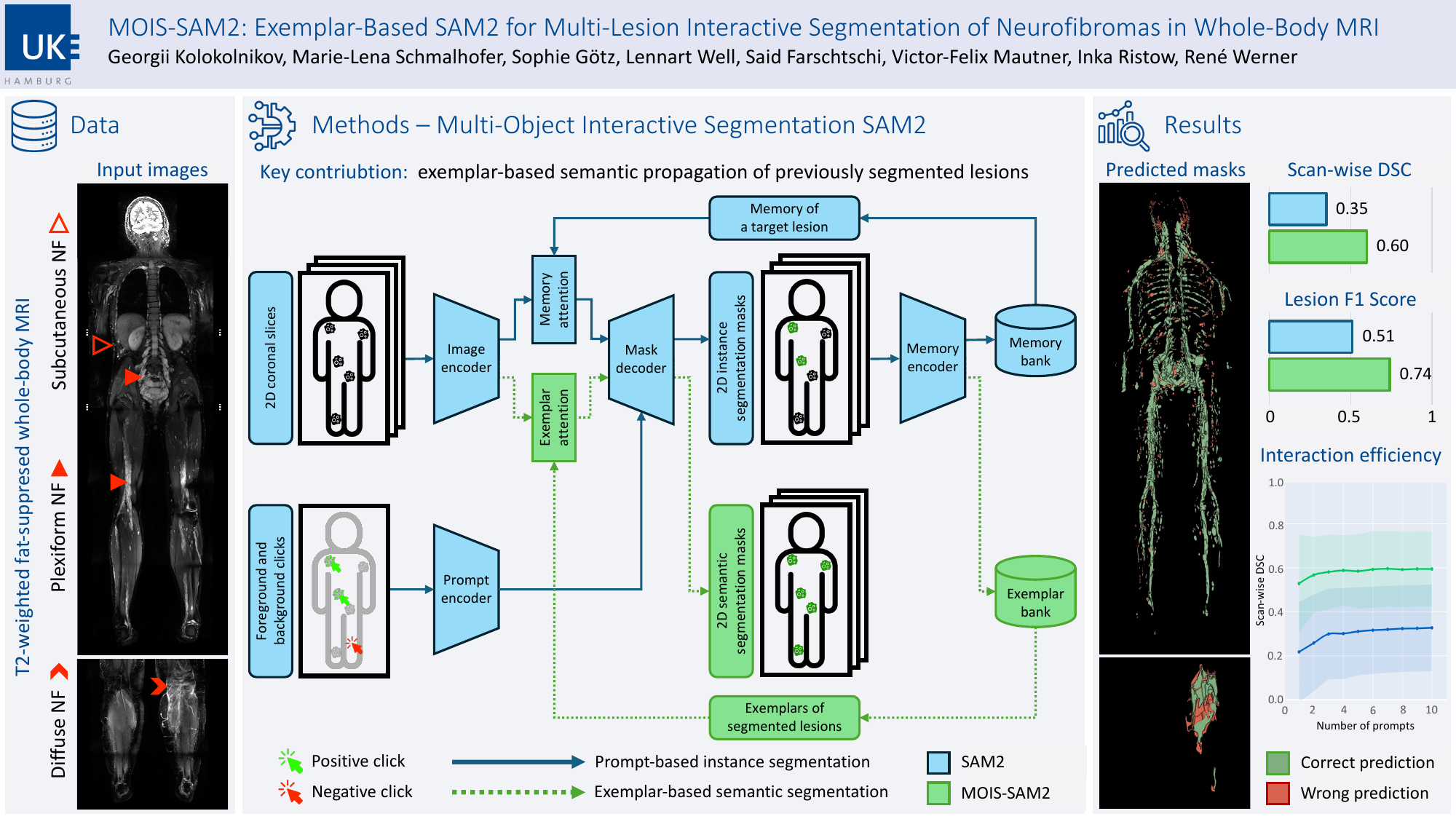}
\end{graphicalabstract}

\begin{highlights}
\item Exemplar-based semantic propagation enables efficient multi-lesion segmentation with substantially reduced manual efforts.
\item The proposed exemplar-based multi-object interactive segmentation model outperforms state-of-the-art convolution- and transformer-based models for neurofibroma segmentation in whole-body MRI.
\item The proposed model achieves expert-level neurofibroma segmentation with strong generalization across domain shift scenarios.
\end{highlights}

%% Keywords

\end{frontmatter}



\section{Introduction}
\label{sec:introduction}

Neurofibromatosis type 1 (NF1) is an autosomal-dominant genetic disorder affecting approximately 1 in 2500 to 3000 individuals worldwide \cite{lammert_prevalence_2005}. A hallmark feature of NF1 is the development of neurofibromas (NFs), benign peripheral nerve sheath tumors, which may be cutaneous, subcutaneous, or plexiform (PNFs) \cite{thakur_multiparametric_2024}. These tumors are highly variable in size, number, and anatomical location, leading to substantial heterogeneity in clinical presentation and disease burden \cite{friedman_type_1997}. Of particular concern are PNFs and distinct nodular lesions (DNLs), both of which carry a risk of transformation into pre-malignant atypical neurofibromatous neoplasm of uncertain biological potential (ANNUBP) or malignant peripheral nerve sheath tumors (MPNSTs), a primary cause of early mortality in NF1 patients \cite{evans_malignant_2002}.

Whole-body magnetic resonance imaging (WB-MRI) with T2-weighted fat-suppressed (T2w) sequences is the clinical standard for longitudinal surveillance of NF1 patients \cite{ahlawat_current_2020}. However, manual segmentation of NFs across the entire body is not only subject to considerable inter- and intra-observer variability but is also time-consuming: depending on the individual tumor burden, segmentation of a single WB-MRI scan may take up to several hours per patient \cite{heffler_tumor_2017, weizman_interactive_2012}. This highlights the need for robust computational solutions.

\begin{figure}
     \centering
     \includegraphics[width=0.8\linewidth]{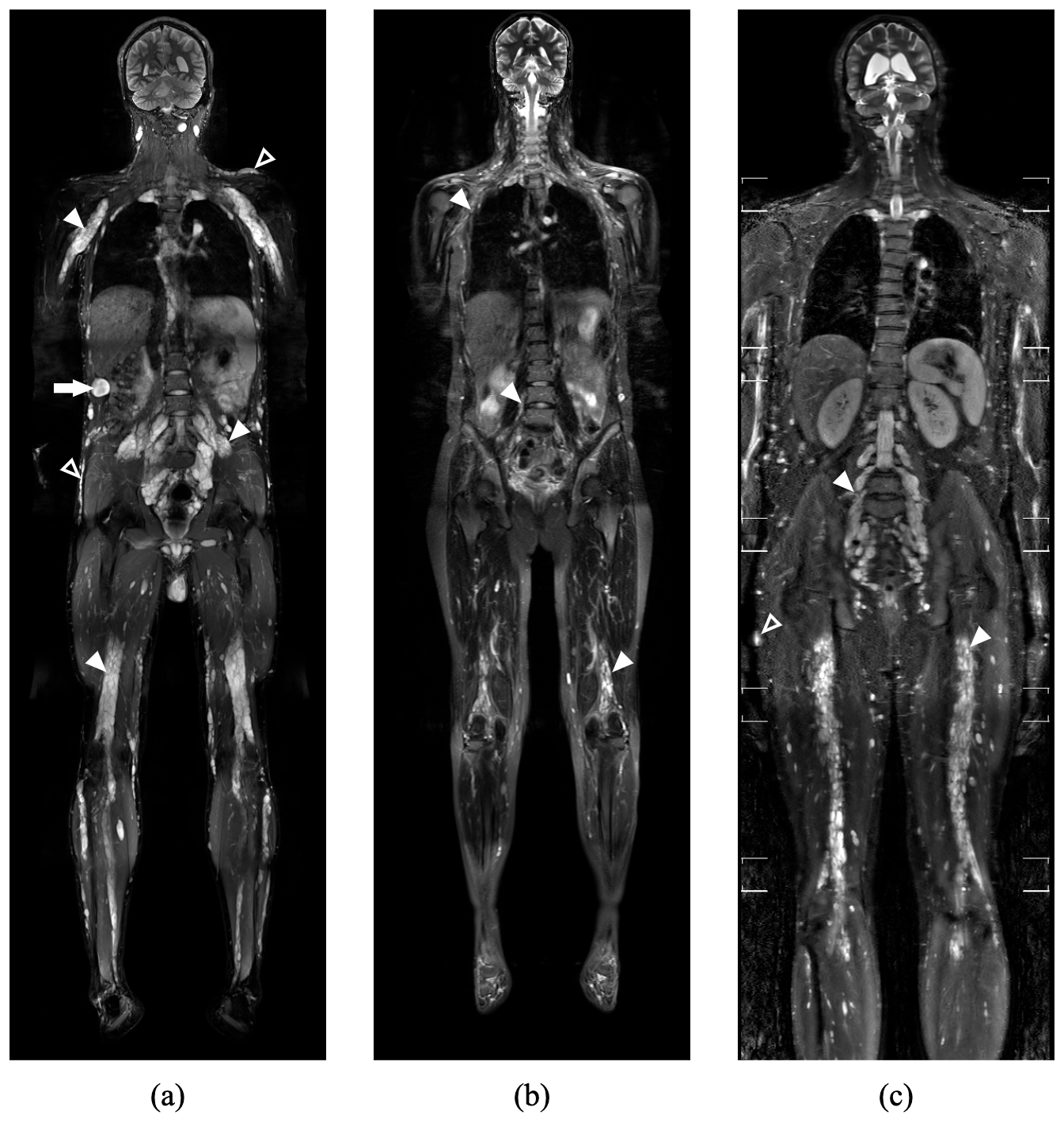}
     \caption{T2-weighted fat-suppressed whole-body MRI scans of three patients with neurofibromatosis type 1, illustrating variability in tumor burden and scanner characteristics. Highlighted examples include plexiform neurofibromas (filled arrowheads), subcutaneous neurofibromas (empty arrowheads), and distinct nodular lesions (filled arrows). (a) A male patient with a tumor burden of 6679 cm$^{3}$, scanned using a 3T Siemens Magnetom scanner. (b) A female patient with a tumor burden of 1316 cm$^{3}$, scanned using a 1.5T Siemens Magnetom scanner. (c) A male patient with a tumor burden of 2348 cm$^{3}$, scanned using a 3T Philips Ingenia scanner.}
     \label{fig:Figure_1}
 \end{figure}

Previous efforts to automate NF segmentation in WB-MRI scans included convolutional neural networks (CNNs), such as nnU-Net \cite{isensee_nnu-net_2021}, which performed well on standard medical tasks but struggled with the high morphological variability of NFs and with domain shifts such as changes in scanner vendor, MRI field strength, and levels of tumor burden in WB-MRI scans (Fig. \ref{fig:Figure_1}). 

Interactive segmentation methods offer greater flexibility by incorporating user feedback \cite{zhang_dins_2022, diaz-pinto_deepedit_2022, hadlich_sliding_2024, isensee_nninteractive_2025}. Convolution-based models like Deep Interactive Networks (DINs) improved NF segmentation but provided only global scan-wise refinement, where a single user interaction influenced the entire scan \cite{zhang_dins_2022}. These models lacked lesion-wise refinement, limiting fine-grained control over individual lesions. More recently, transformer-based models like the Segment Anything Model 2 (SAM2) demonstrated better generalization and control over individual instance refinement \cite{liu_simpleclick_2023, kirillov_segment_2023, ravi_sam_2024, he_vista3d_2024}. However, their scalability to multi-object segmentation scenarios remains unexplored. SAM2 requires a separate user click for each individual lesion, which becomes infeasible for NF1 patients with hundreds of lesions in a single WB-MRI scan \cite{heffler_tumor_2017}.

These limitations highlight the need for interactive segmentation methods that can efficiently handle large numbers of lesions in WB-MRI scans of NF1 patients while enabling precise lesion-wise refinement. We address this gap by introducing MOIS-SAM2, a multi-object interactive segmentation extension of SAM2 tailored to NF1. Our model leverages exemplar-based semantic propagation: previously segmented lesions are stored as exemplars in an exemplar bank and used to guide segmentation of non-prompted lesions via an exemplar attention. This approach combines the precision of instance-wise refinement with the scalability required for high tumor-burden cases. We hypothesize that exemplar attention enhances NF segmentation accuracy while reducing the manual interaction burden compared to existing interactive methods \cite{zhang_dins_2022, ravi_sam_2024, he_vista3d_2024}. 

Key contributions of our work are:

\begin{itemize}
    \item A novel segmentation model that extends SAM2 with exemplar attention and exemplar bank, enabling multi-lesion interactive segmentation of NFs in WB-MRI scans.
    \item A lesion-wise evaluation pipeline to assess interactive segmentation performance in multi-lesion settings.
    \item Publicly available code, including training scripts, the lesion-wise evaluation pipeline, and full integration with 3D Slicer \cite{fedorov_3d_2012} via MONAI Label \cite{diaz-pinto_monai_2024}, enabling reproducibility and deployment.
\end{itemize}

Together, these contributions represent a step toward more efficient segmentation of NFs in WB-MRI scans.

\section{Related Work}
\label{sec:related_work}

\subsection{Segmentation of Neurofibromas in WB-MRI}
\label{subsec:segmentation_of_neurofibromas_in_wb-mri}
Early approaches to NF segmentation in WB-MRI scans primarily relied on semi-automated methods. Early semi-automated methods combined region growing with edge detection or dynamic-threshold level sets, achieving reproducible results but requiring manual ROI placement and struggling with iso-intense and low-contrast regions \cite{solomon_automated_2004, cai_tumor_2009}. Weizman \textit{et al.} later introduced interactive tools based on histogram modeling and 3D region growing, reducing annotation time by up to 80\% \cite{weizman_pnist_2014}. However, these methods relied on 2D slice-wise initialization and manual refinement.

Subsequent studies adopted CNNs for NF segmentation. Ho \textit{et al.} used multi-spectral model to improve lesion characterization \cite{ho_image_2020}, while Wu \textit{et al.} introduced hybrid models combining CNNs with active contours (DPAC, DH-GAC) to enhance boundary accuracy, though at high computational cost and limited to 2D \cite{wu_deep_2020, wu_dh-gac_2022}. Zhang \textit{et al.} proposed DINs, the first interactive deep learning model for NF segmentation, embedding user clicks via Exponential Distance Transforms into a 3D anisotropic U-Net \cite{zhang_dins_2022}. While DINs improved over nnU-Net and classical methods, the model propagated each user click globally across the entire scan, affecting the segmentation of all lesions simultaneously and limiting the ability to selectively refine individual lesions with high precision. A recent semi-automated approach by Kiaei \textit{et al.} applied mean intensity thresholding across MRI sequences but still required manual ROI input \cite{kiaei_development_2024}.

Recent efforts explored automated NF segmentation pipelines that incorporated anatomical knowledge and used cascaded deep learning models. In our previous conference contribution \cite{kolokolnikov2024enhancing}, we introduced an anatomy-informed multi-stage pipeline using TotalSegmentator \cite{wasserthal_totalsegmentator_2023} to divide the body into anatomical zones, and dedicated U-Net models for each zone to segment NFs. Similarly, Wei \textit{et al.} proposed two deep learning models, a cascaded U-Net+ResNet18 and a YOLOv5-based architecture \cite{wei_multicenter_2025}. These works benefited from incorporating anatomical context, underscoring  the importance of organ-aware segmentation in NF1 patients. 

\subsection{Interactive Segmentation Methods}
\label{subsec:interactive_segmentation_methods}

As demonstrated by Zhang \textit{et al.} \cite{zhang_dins_2022}, deep learning-based interactive segmentation methods offer a promising alternative to automated models for complex segmentation tasks such as NF segmentation.

Before the wide application of deep learning methods, classical approaches laid the foundation for interactive segmentation. GrabCut, Random Walks, and Graph Cuts \cite{rother_grabcut_2004, grady_random_2006, boykov_interactive_2001} used minimal input (e.g., bounding boxes, seeds) to iteratively refine segmentations. While effective in 2D, they lacked scalability to 3D and robustness for medical imaging.

With the advent of deep learning, more powerful frameworks emerged. Deep Interactive Object Selection introduced positive/negative clicks as distance maps for precise 2D segmentation \cite{xu_deep_2016}, and DEXTR used extreme points to generate high-accuracy object masks \cite{maninis_deep_2018}. For medical imaging, CNN-based models were adapted to volumetric data. DeepEdit combined non-interactive segmentation methods, such as nnU-Net, U-Net, or UNETR, with the interactive segmentation method DeepGrow \cite{diaz-pinto_deepedit_2022}. SW-FastEdit improved scalability of DeepEdit via sliding window inference and patch-wise click correction for full-body positron emission tomography (PET) scans \cite{hadlich_sliding_2024}. More recently, nnInteractive extended nnU-Net for few-shot multi-label segmentation with one-click-per-class inference \cite{isensee_nninteractive_2025}. However, most CNN-based approaches applied user prompts at the scan level, lacking the ability to localize interactions to individual lesions. As a result, they offered limited support for fine-grained, instance-wise refinement.

Recent transformer-based frameworks marked a shift toward general-purpose, prompt-driven segmentation. Faizov \textit{et al.} proposed a click-token SegFormer-based model for iterative segmentation refinement \cite{faizov_interactive_2022}. iSegFormer extended transformer-based models to 3D medical images via slice-wise interaction and segmentation propagation, demonstrating high accuracy on knee MRI \cite{wang_isegformer_2022}. SimpleClick adopted a plain Vision Transformer (ViT) architecture, achieving strong results in both natural and medical data \cite{liu_simpleclick_2023}. The Segment Anything Model (SAM) introduced a universal promptable segmentation model trained on over a billion masks \cite{kirillov_segment_2023}. However, its 2D design limited its applicability to medical images. SAM-Med3D adapted SAM to volumetric medical data and integrated 3D positional encodings \cite{wang2023sam}, which led to a better performance in 3D contexts.

State-of-the-art transformer-based interactive segmentation methods are primarily built on the SAM foundation. Ravi \textit{et al.} introduced SAM2 \cite{ravi_sam_2024} that featured a streaming transformer architecture with a memory attention module. MedSAM-2 further extended SAM2 to treat medical volumes as video sequences, enabling one-prompt segmentation across an entire 3D MRI or CT scan \cite{zhu2024medical}. The Versatile Imaging Segmentation and Annotation Model (VISTA3D) represents the current state-of-the-art in 3D transformer-based interactive segmentation \cite{he_vista3d_2024}. It uses a SwinUNETR-style backbone with prompt-aware decoding, integrating 2D and 3D user inputs into the attention mechanism. While VISTA3D achieves zero-shot generalization and high consistency, its high computational cost limits real-time usability.

\subsection{Multi-Lesion Interactive Segmentation}
\label{subsec:multi-lesion_interactive_segmentation}

Existing interactive segmentation models face a fundamental trade-off between scalability and precision. Convolution-based models can propagate user input across an entire scan by leveraging their receptive fields and spatial priors \cite{chen_transunet_2024}, enabling multi-lesion segmentation with few prompts – but at the cost of precise lesion-wise refinement. Transformer-based models, in contrast, support localized instance-wise segmentation through attention mechanisms \cite{chen_transunet_2024}, but typically require a separate prompt per instance, limiting their scalability. In WB-MRI of NF1 patients, where hundreds of lesions may be present \cite{heffler_tumor_2017}, neither approach alone is sufficient. This motivates the need for multi-object interactive segmentation: models that enable efficient multi-lesion segmentation while retaining the ability to refine individual instances with minimal user input.

Recent methods began addressing the problem of multi-object interactive segmentation. The Dynamic Multi-Object Interactive Segmentation Transformer (DynaMITe) introduced spatio-temporal transformer queries for joint multi-object segmentation in a single forward pass \cite{rana_dynamite_2023}. Li \textit{et al.} introduced iCMFormer++ \cite{li2024learning}, an exemplar-based approach in which previously segmented lesions acted as exemplars to guide the segmentation of similar objects. A two-stream transformer with cross-attention aligned exemplar features with target images, enhanced by a similarity-driven exemplar-informed module, enabling generalization across similar objects with fewer interactions. However, the exemplar retrieval relied on convolutional similarity matching within the image space, limiting its ability to capture more abstract semantic relationships and potentially reducing robustness in cases of large intra-class variability. Wu \textit{et al.} introduced the One-Prompt Segmentation paradigm, which bridged the gap between interactive and one-shot learning in medical image segmentation \cite{wu_one-prompt_2024}. Rather than prompting each target individually or requiring multiple labeled examples, this method leveraged a single prompted template image to guide segmentation on a new, unseen task in a single forward pass.
	
These innovations collectively advanced the field of multi-object interactive segmentation. However, none yet fully resolved the demands posed by NF1 cases: dense, same-class lesion distributions requiring high precision, annotation consistency, and low interaction burden, while also enabling both single- and multi-lesion segmentation.

\section{Methods}
\label{sec:methods}

We address the challenge of interactive segmentation of numerous, morphologically diverse, and spatially scattered NFs in T2w WB-MRI. To this end, we propose MOIS-SAM2 – a multi-object interactive segmentation extension of SAM2 – designed to enable exemplar-based multi-lesion segmentation of NFs with minimal user input.

\subsection{Problem Formulation}
\label{subsec:problem_formulation}

Let $\mathcal{I} \in \mathbb{R}^{H \times W \times D}$ denote a 3D WB-MRI scan composed of $D$ 2D slices, where $H$ and $W$ are the height and width. Each scan may contain up to $N$ NF lesions $\{L_n\}_{n=1}^{N}$, all belonging to the same semantic class. Throughout this work, we refer to each coronal image slice as a 2D slice (the term 'slice-level'), and use the term 'scan-level' to refer to the full 3D volume. Accordingly, segmentation masks can be either slice-level (2D) or scan-level (3D), depending on the context.

A user interaction is defined as a click $c = (x, y, d, l)$, where $(x, y)$ is a 2D spatial coordinate on slice $d \in [0, D{-}1]$, and $l \in \{0, 1\}$ indicates background ($0$) or foreground ($1$) content. A click set is denoted by $\mathcal{C} = \{c_m\}_{m=1}^{M}$.

When a lesion \( L_n \) is segmented given a click set \( \mathcal{C} \), its semantic representation is stored as an exemplar \( e_k \), where \( k \in \{1, \dots, K\} \) indexes the \( K \) total exemplars stored in the exemplar bank. Each exemplar \( e_k \) is associated with a 2D slice-level single-lesion instance segmentation mask \( \hat{m}_n^{(d)} \in \{0,1\}^{H \times W} \) on slice \( d \), and consists of the following components:

\begin{itemize}
    \item Visual embedding ${z}_k \in \mathbb{R}^{d_{\text{vis}}}$: a learned feature vector obtained by summing a downsampled version of the slice-level single-lesion segmentation mask $\hat{m}_n^{(d)}$ with the unconditioned image encoder features of a WB-MRI slice $d$, followed by light-weight convolutions, as in SAM2.
    \item Positional encoding ${p}_k \in \mathbb{R}^{d_{\text{pos}}}$: a fixed sinusoidal spatial embedding that encodes the intra-slice location of the segmented lesion $L_n$.
    \item Object slice index $d_k \in [0, D{-}1]$: the index of the slice on which the lesion $L_n$ was segmented.
    \item Prompt flag $flag_k \in \{0, 1\}$: indicates whether the exemplar was acquired using a user click or inferred automatically via memory-based slice-wise propagation of segmentation as in SAM2.
    \item Object pointer \( {v}_k \in \mathbb{R}^{d_{\text{ptr}}} \): an embedding vector derived from the output token of the mask decoder corresponding to the segmented lesion \( L_n \), following the SAM2 mechanism. This vector captures object-level information and is reused across all \( D \) slices of the WB-MRI scan \( \mathcal{I} \) to maintain consistent lesion identity during slice-wise segmentation propagation.

\end{itemize}

MOIS-SAM2 is a multi-object interactive segmentation model, represented as a function $f_\theta(\mathcal{I}, \mathcal{C}, \mathcal{E})$ parameterized by weights $\theta$. It takes as input a WB-MRI scan $\mathcal{I}$, a click set $\mathcal{C}$, and a set of exemplars $\mathcal{E} = \{e_k\}_{k=1}^{K}$, and supports two inference modes:

\begin{itemize}
    \item Prompt-based instance segmentation (as in SAM2): given a set of user clicks $\mathcal{C}$, the model produces a binary scan-level single-lesion instance segmentation mask:    
    \begin{equation}
        \hat{M}_{\text{single-lesion}} = f_\theta(\mathcal{I}, \mathcal{C}, \emptyset) \in \{0,1\}^{H \times W \times D},
    \end{equation}
    where only the lesion corresponding to the click is labeled as foreground.

    \item {Exemplar-based semantic segmentation}: given a set of exemplars $\mathcal{E}$ derived from previously segmented lesions, the model predicts a binary scan-level multi-lesion semantic segmentation mask
    
    \begin{equation}
        \hat{M}_{\text{multi-lesion}} = f_\theta(\mathcal{I}, \emptyset, \mathcal{E}) \in \{0,1\}^{H \times W \times D},
    \end{equation}
    where all semantically similar lesions across the WB-MRI scan are labeled as foreground. Here, semantic similarity refers to lesions whose visual embeddings – derived from both the image and exemplar mask – are close in the learned feature space, as determined by the exemplar attention module.

\end{itemize}

\subsection{Model Architecture}
\label{subsec:model_architecture}

MOIS-SAM2 builds on SAM2 \cite{ravi_sam_2024} and its medical adaptation MedSAM2 \cite{zhu2024medical}, which processes 3D medical images as pseudo-video sequences of 2D slices. This design choice is particularly suitable for our use case, as WB-MRI scans are highly anisotropic,  with a high in-plane resolution and a relatively small number of coronal slices. Processing these volumes as 2D sequences is both computationally efficient. In contrast, fully 3D transformer-based methods such as VISTA3D \cite{he_vista3d_2024} require substantially more memory and computation.

\begin{figure}[t]
    \centering
    \includegraphics[width=1.0\linewidth]{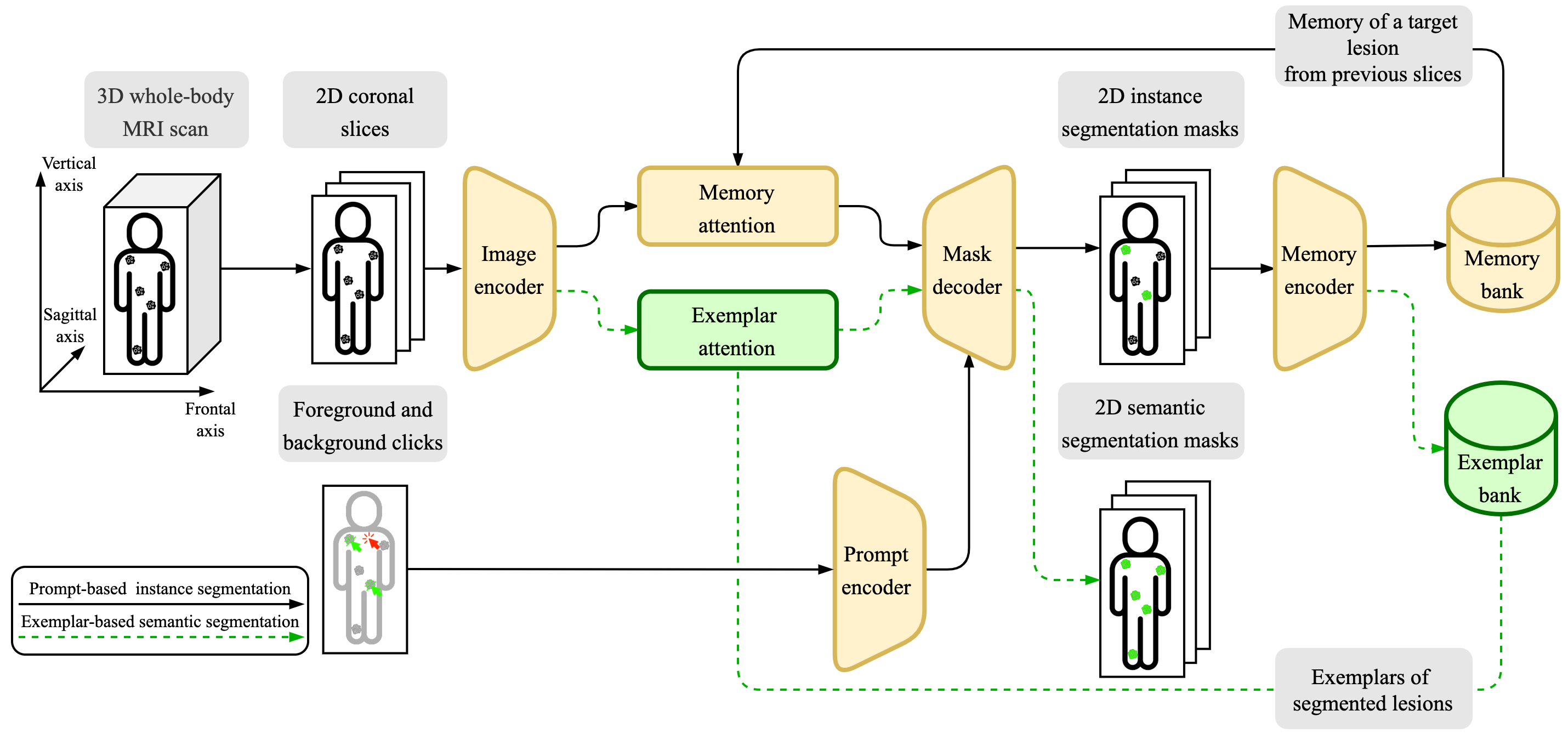}
    \caption{Overview of the proposed model architecture MOIS-SAM2 for multi-lesion interactive segmentation of neurofibromas in whole-body MRI scans. The model processes coronal slices from a whole-body MRI scan and combines two segmentation approaches: (1) Prompt-based instance segmentation (solid arrows), where user clicks (positive - green, negative - red) guide slice-level single-lesion segmentation, and the resulting mask is propagated across slices using the memory bank and memory attention; (2) Exemplar-based semantic segmentation (dashed green arrows), where previously segmented lesions are stored in the exemplar bank and retrieved via exemplar attention to perform exemplar-based semantic propagation to unprompted lesions. The architecture combines base modules from Segment Anything Model 2 (SAM2) \cite{ravi_sam_2024} (orange) with MOIS-SAM2-specific extensions (green), enabling both lesion-wise refinement and scalable multi-lesion segmentation.}
    \label{fig:Figure_2}
\end{figure}

Our model preserves the core functionality of SAM2 \cite{ravi_sam_2024} – prompt-based instance segmentation and memory-based slice-wise propagation – while introducing a novel exemplar-based semantic propagation module for multi-lesion segmentation. This module is inspired by the exemplar-based propagation paradigm proposed by Li \textit{et al.} \cite{li2024learning}, but unlike that method, which matches a single exemplar via convolutional similarity in image space, our model performs cross-attention-based matching across multiple exemplars simultaneously in the learned feature space, enabling more abstract semantic alignment. Moreover, we build directly on the state-of-the-art SAM2 architecture, ensuring strong generalization and robust memory propagation over slices in 3D. An overview of the MOIS-SAM2 architecture is shown in Figure~\ref{fig:Figure_2}.

The core components inherited from SAM2 include:
\begin{itemize}
    \item Image encoder: extracts dense, multi-scale visual embeddings for each slice.
    \item Prompt encoder: encodes user clicks as spatially aware tokens.
    \item Memory encoder: encodes slice-level single-lesion segmentation masks and corresponding vision embeddings from the image encoder into compact visual embeddings.
    \item Memory bank: stores recent slice-level single-lesion segmentation masks for consistency of memory-based slice-wise segmentation propagation.
    \item Memory attention: integrates relevant context from the memory bank into the next slice visual embeddings to provide memory-based slice-wise propagation of segmentation masks.
    \item Mask decoder: integrates visual embeddings and prompts to predict slice-level single-lesion segmentation masks.
\end{itemize}

To enable multi-object segmentation of non-prompted lesions by exemplar-based semantic propagation, we introduced the following key modules:

\begin{itemize}
    \item Exemplar bank: a dynamic memory structure that stores exemplars of previously segmented lesions.
    \item Exemplar attention: an attention mechanism with rotary positional encoding, structurally identical to the memory attention in SAM2. It is functionally specialized to perform cross-attention between current slice visual embeddings and the exemplar bank, enabling the detection of multiple semantically similar lesions without user clicks.
\end{itemize}

While the exemplar attention reuses the architectural design of memory attention from SAM2, it is trained for a distinct purpose: exemplar attention generalizes segmentation to novel lesions across the scan, whereas memory attention propagates a known lesion across slices. As shown in our ablation study (Section \ref{subsec:architecture_ablation_results}), attempting to use a shared attention module for both tasks leads to a degraded performance – motivating the need for dedicated attention pathways.

To ensure memory efficiency of MOIS-SAM2, the exemplar bank is limited to $K$ entries and follows the prioritized replacement strategy, when this limit is exceeded:
\begin{itemize}
    \item Prompted-over-non-prompted priority: prompted exemplars replace non-prompted ones. 
    \item New-over-old priority: among exemplars with the same prompt flag, more recently added exemplars replace older ones. This favors fresh context within the same category (prompted or non-prompted).
    \item Close-over-distant priority: exemplars are sorted by inter-slice distance to the current ROI, prioritizing spatially relevant context.
\end{itemize}

When a previously segmented lesion is refined with new user clicks, its corresponding exemplar is updated accordingly. If the exemplar was originally inferred without user input, its $flag_k$ is switched to indicate a prompted exemplar.

\subsection{Inference Logic}
\label{subsec:inference_logic}

The full inference pipeline of MOIS-SAM2 comprises two stages: 

\begin{enumerate}
    \item Prompt-based single-lesion instance segmentation with memory-based slice-wise propagation, and
    \item Exemplar-based multi-lesion semantic segmentation of the remaining unprompted lesions.
\end{enumerate}
 
The goal of the full inference pipeline is to predict a scan-level multi-lesion segmentation mask $\hat{M}_{\text{multi-lesion}}$ that covers all NFs in the WB-MRI scan with minimal user input. Algorithm~\ref{alg:mois_seg} summarizes the complete inference process.

\begin{figure}[ht]
\scriptsize % or \footnotesize or \scriptsize
\begin{algorithm}[H]
\caption{Full inference pipeline of MOIS-SAM2}
\label{alg:mois_seg}
\KwIn{WB-MRI scan $\mathcal{I}$, click set $\mathcal{C}$}
\KwOut{Scan-level multi-lesion segmentation mask $\hat{M}_{\text{multi-lesion}} \in \{0,1\}^{H \times W \times D}$}

Initialize memory bank $\mathcal{M} \leftarrow \emptyset$, exemplar bank $\mathcal{E} \leftarrow \emptyset$\;

\tcc{\textbf{Stage 1: Prompt-based single-lesion instance segmentation}}
\ForEach{slice $d$ in $\mathcal{I}$}{
    Extract slice-level visual embedding ${z}^{(d)}$ using the image encoder\;
    
    \If{click subset $\mathcal{C}_d \neq \emptyset$}{
        Predict slice-level single-lesion segmentation mask $\hat{m}_n^{(d)}$ based on click subset $\mathcal{C}_d$ and the visual embedding ${z}^{(d)}$\;
        Update memory bank $\mathcal{M}$ with $\hat{m}_n^{(d)}$\;
        Update exemplar bank $\mathcal{E}$ with the corresponding prompted exemplar $e_k^{(d)}$\;
    }
    \ElseIf{memory bank $\mathcal{M} \neq \emptyset$}{
        Predict slice-level single-lesion segmentation mask $\hat{m}_n^{(d)}$ based on memory bank $\mathcal{M}$ and the visual embedding ${z}^{(d)}$\;
        Update memory bank $\mathcal{M}$ with $\hat{m}_n^{(d)}$\;
        Update exemplar bank $\mathcal{E}$ with the corresponding non-prompted exemplar $e_k^{(d)}$\;
    }
    \Else{
    Skip slice $d$ (no prompt and no memory)\;
}
}

\tcc{\textbf{Stage 2: Exemplar-based multi-lesion semantic segmentation}}
\ForEach{slice $d$ in $\mathcal{I}$}{
    Extract slice-level visual embedding ${z}^{(d)}$ using the image encoder\;
    Apply exemplar attention between ${z}^{(d)}$ and exemplar bank $\mathcal{E}$:  $\tilde{{z}}_d = \text{CrossAttn}({z}^{(d)}, \mathcal{E})$\;
    Predict slice-level multi-lesion segmentation semantic mask $\hat{m}_{\text{multi-lesion}}^{(d)}$ based on cross-attended embedding $\tilde{{z}}_d$\;
    Add $\hat{m}_{\text{multi-lesion}}^{(d)}$ to the scan-level multi-lesion segmentation mask $\hat{M}_{\text{multi-lesion}}$\;
}
\Return{$\hat{M}_{\text{multi-lesion}}$}
\end{algorithm}
\end{figure}

During the exemplar-based inference stage, each slice is processed independently. The visual embedding of the current slice ${z}^{(d)}$ is combined with the visual embeddings from the exemplar bank $\mathcal{E}$ using the cross-attention mechanism of the exemplar attention module. This results in a cross-attended embedding $\tilde{{z}}_d$. This step aims to identify structures in the current slice that are semantically similar to previously segmented lesion stored in the exemplar bank.

The cross-attended embedding is passed to the mask decoder, which predicts the binary slice-level multi-lesion segmentation mask $\hat{m}_{\text{multi-lesion}}^{(d)} \in \{0,1\}^{H \times W}$. If no exemplars are available, a learned no-exemplar token is used to ensure robust fallback behavior. The slice-level multi-lesion segmentation masks are then merged into the scan-level multi-lesion segmentation mask $\hat{M}_{\text{multi-lesion}}$.

The final scan-level multi-lesion segmentation mask $\hat{M}_{\text{multi-lesion}}$ is refined using optional morphological operations after applying connected component analysis:

\begin{itemize}
    \item Small component removal: removes isolated artefacts based on a minimum volume threshold $V_{\text{thresh}}$.
    \item Hole filling: applied to each connected component in the binary mask to improve the structural continuity of lesions.
\end{itemize}

This two-stage inference pipeline allows MOIS-SAM2 to scale from sparse, prompt-based single-lesion instance segmentation to exemplar-based multi-lesion semantic segmentation in WB-MRI scans.

\subsection{Training Strategy}
\label{subsec:training_strategy}

MOIS-SAM2 is trained to jointly support prompt-based single-lesion instance segmentation and exemplar-based multi-lesion semantic segmentation. The model is optimized using a multi-task supervision strategy, with all supervision and loss computation performed at the 2D slice level. This design follows the SAM2 training paradigm and reflects the 2D-slice-based inference algorithm, allowing efficient memory usage.

Each training sample consists of:
\begin{itemize}
    \item A short sequence of consecutive coronal slices $\{I_i\}_{i=0}^{D_{train}-1}$ from a WB-MRI scan $\mathcal{I}$, where $D_{train} < D$. The number of slices $D_{\text{train}}$ per sequence is chosen based on available GPU memory constraints
    \item A subset of prompted lesions $\{L_j\}_{j=0}^{N_{train}}$ (i.e., lesions selected to receive simulated user clicks during training) within the sequence of consecutive coronal slices, where $N_{train} < N$.
    \item A respective sequence of 2D coronal slice-level ground truth masks:
    \begin{itemize}
        \item Instance masks $\{\{m_j^{(i)}\}_{j=0}^{N_{train}}\}_{i=0}^{D_{train}-1}$ of individual prompted lesions $L_j$; used in the prompt-based supervision.
        \item Semantic masks $\{{m}_{\text{multi-lesion}}^{(i)}\}_{i=0}^{D_{train}-1}$ of all lesions $\{L_n\}_{n=1}^{N}$ visible on slice $i$; used in the exemplar-based supervision.
    \end{itemize}
\end{itemize}

Initial clicks are simulated by placing a foreground click at the 3D geometric center of a scan-level single-lesion masks. Additional correction clicks are placed at the center of the discrepancy between the predicted and ground truth masks, simulating iterative user interaction. 

The model is optimized using a composite multi-task loss function. In addition to the original instance segmentation loss $\mathcal{L}_{\text{instance}}$ and object presence loss $\mathcal{L}_{\text{object}}$ used in SAM2 \cite{ravi_sam_2024}, we introduce a semantic segmentation loss, calculated as the average sum of the binary cross-entropy, Dice, and Intersection over Union (IoU) losses between the predicted and ground truth slice-level multi-lesion segmentation masks over the $D_{\text{train}}$ coronal slices. The total loss is defined as the sum of all loss terms.

\section{Experiments}
\label{sec:experiments}

Our work presents a retrospective study on interactive NF segmentation in fat-suppressed T2w WB-MRI scans of patients with NF1. It adheres to the Checklist for Artificial Intelligence in Medical Imaging (CLAIM) reporting guidelines \cite{tejani_checklist_2024}.

\subsection{Data Description}
\label{subsec:data_description}

\subsubsection{Data Characteristics}
\label{subsubsec:data_characteristics}

This retrospective study was built on a WB-MRI dataset from a patient cohort at the University Medical Center Hamburg-Eppendorf (UKE). The dataset comprised 119 T2w WB-MRI scans from 84 unique NF1 patients, collected between 2006 and 2024 at two clinical sites (UKE and Altona), and acquired using fat-suppressed T2w coronal sequences. Inclusion criteria required patients to (1) meet the diagnostic criteria for NF1 from the National Institutes of Health (NIH) \cite{noauthor_neurofibromatosis_1988}; and (2) have undergone T2w WB-MRI with visible peripheral nerve sheath tumors acquired using a standardized coronal protocol.

Exclusion criteria were therapy-induced confounders such as resected tumors, severe motion artefacts, or incomplete imaging coverage. Data collection was approved by the local ethics committee (2022-300201-WF, 2022-300201\_1-WF, and 2022-300201\_2-WF), in compliance with data protection regulations and the Declaration of Helsinki.

Due to the sensitive nature of the WB-MRI data – which included full-body coverage with identifiable features such as the head – along with the rarity of NF1 and the distinctiveness of NF distribution patterns, public data release poses a risk of patient re-identification. Consequently, data access is limited in compliance with institutional policies and ethical regulations.

The dataset was divided into a training/validation set and four independent test sets with no patient overlap. Tumor burden categories were defined based on the total tumor volume across the whole dataset:
\begin{itemize}
    \item Low tumor burden: tumor volume falling within the first quartile (tumor volume $<$ 50\,cm\textsuperscript{3}).
    \item High tumor burden: tumor volume falling within the third quartile (tumor volume $>$ 3000\,cm\textsuperscript{3}).
    \item Medium tumor burden: cases between the first and third quartiles.
\end{itemize}

\noindent The data subsets were defined as follows:

 \begin{itemize}
     \item Training Set (n = 63 scans / 42 patients): acquired at UKE with a Siemens 3T scanner (Magnetom), containing medium to high tumor burden. Used for 3-fold cross-validation, with partitioning at the patient level to ensure disjoint subsets.
     \item Test Set 1 (n = 13 scans / 13 patients): in-domain benchmark acquired at UKE with a 3T Siemens scanner. Matched to the training set in terms of tumor and imaging characteristics.
     \item Test Set 2 (n = 11 scans / 9 patients): domain shift benchmark acquired at UKE with a 1.5T Siemens scanner. Used to assess models robustness to MRI field strength variation.
     \item Test Set 3 (n = 22 scans / 10 patients): subset of low tumor burden cases acquired at UKE using a 3T Siemens scanner. Used to assess model sensitivity in sparse lesion scenarios.
     \item Test Set 4 (n = 10 scans / 10 patients): external subset acquired at the Altona clinical site with a Philips 3T scanner (Ingenia), containing medium to high tumor burden. Used to assess models performance under domain shift in both clinical site and scanner vendor.
 \end{itemize}

A demographic and imaging summary of the dataset is presented in Table~\ref{tab:data_description}. Across all sets, scans showed high spatial anisotropy with the median spacing 0.62 mm $\times$ 0.62 mm $\times$ 7.80 mm and included between 20 to 51 coronal slices. Median tumor volumes and counts varied widely, reflecting high inter-patient heterogeneity typical for NF1.

\begin{table}[t]
\centering
\scriptsize
\renewcommand{\arraystretch}{1.3}
\begin{tabular}{>{\raggedright\arraybackslash}p{0.18\linewidth}>{\centering\arraybackslash}p{0.12\linewidth}>{\centering\arraybackslash}p{0.12\linewidth}>{\centering\arraybackslash}p{0.13\linewidth}>{\centering\arraybackslash}p{0.12\linewidth}>{\centering\arraybackslash}p{0.12\linewidth}}
\hline
\textbf{Characteristic} & \textbf{Training} & \textbf{Test Set 1} & \textbf{Test Set 2} & \textbf{Test Set 3} & \textbf{Test Set 4} \\
\hline
Scans (n) & 63 & 13 & 11 & 22 & 10 \\
Patients (n) & 42 & 13 & 9 & 10 & 10 \\
Clinical site & Altona & Altona & Altona & Altona & UKE \\
Scanner & Siemens Magnetom 3T & Siemens Magnetom 3T & Siemens Magnetom 1.5T & Siemens Magnetom 3T & Philips Ingenia 3T \\
Sex (m/f) & 21 / 21 & 8 / 5 & 3 / 6 & 5 / 5 & 4 / 6 \\
Age (mean ± SD, y) & 31.9 ± 13.4 & 32.5 ± 13.6 & 27.7 ± 13.6 & 22.4 ± 15.7 & 34.1 ± 12.2\\
Time span & 2012–2021 & 2013–2020 & 2006–2012 & 2013–2021 & 2021–2024 \\
Slices & 28–34 & 27–32 & 20–31 & 22–33 & 42–51 \\
Tumor volume (Med, IQR, cm$^3$) & 638.7, 1296.9 & 244.0, 313.5 & 438.1, 408.1 & 6.0, 9.6 & 215.9, 1105.2 \\
Tumor count (Med, IQR) & 244, 313 & 244, 313 & 112, 272 & 2, 5 & 151, 333. \\
\hline
\multicolumn{6}{p{0.9\linewidth}}{SD – standard deviation, Med – median, IQR – inter-quartile range, m – male, f – female.}\\  

\end{tabular}
\caption{Summary of the dataset characteristics used for training, validation, and testing. Test sets correspond to different domain shift scenarios: in-domain (Test Set 1); MRI field strength variation (Test Set 2); low tumor burden (Test Set 3); clinical site and scanner vendor variation (Test Set 4).}
\label{tab:data_description}
\end{table}

\subsubsection{Data Annotation}
\label{subsubsec:data_annotation}
Ground-truth was acquired through manual tumor annotation performed by two radiologists (I.R., M.-L.S.). Annotation guidelines defined NFs as hyperintense lesions relative to surrounding tissue. Annotations excluded cutaneous NFs and were limited to internal NFs, PNFs and DNLs due to the elevated risk of malignant transformation. Tumor boundaries were manually contoured slice by slice to create volumetric 3D ROIs. Each annotation result was saved as a ground truth binary multi-lesion mask. 

Manual expert annotation was chosen as the reference standard to avoid the bias of existing automated or semi-automated methods when segmenting NFs in WB-MRI across the full anatomical range and lesion variability encountered in NF1.  While manual delineation is considered the clinical gold standard, it is subject to inter-observer variability. To mitigate this and ensure anatomical consistency, axial planes were consulted during annotation, and representative cases were jointly reviewed by the annotators. 

\subsubsection{Data Preprocessing}
\label{subsubsec:data_preprocessing}
Prior to training and inference of the MOIS-SAM2 model, all data were preprocessed as follows:
 \begin{enumerate}
     \item Spacing normalization: all scans were resampled to the median spacing of 0.62~mm $\times$ 0.62~mm $\times$ 7.80~mm to ensure uniform voxel resolution.
     \item Intensity normalization: percentile-based normalization (0.5 and 99.5 percentiles) was applied to each scan to mitigate inter-scan contrast variability.
     \item Slice extraction: scans were split into individual 2D slices along the anterior–posterior axis to align with the SAM2 pseudo-video sequence processing concept.
     \item Spatial resizing: each slice was resized to 1024×1024 pixels to align with the input resolution used by the SAM2 model.
 \end{enumerate}
 
Annotation masks were additionally processed for training purposes to create both:
 \begin{itemize}
     \item Instance segmentation masks, where connected components defined individual lesions for supervision of  the prompt-based single-lesion segmentation.
     \item Semantic segmentation masks representing all lesions for supervision of the exemplar-based multi-lesion segmentation.
 \end{itemize}

\subsection{Experimental Design}
\label{subsec:experimental_design}

\subsubsection{Evaluation Pipeline}
\label{subsubsec:evaluation_pipeline}

Interactive segmentation of NFs in WB-MRI can be evaluated at two levels: scan-wise and lesion-wise. In the scan-wise scenario, which is commonly used in the literature \cite{marinov_deep_2024}, simulated user clicks target the region of the largest segmentation error across the entire scan. While this can effectively improve overall semantic segmentation accuracy, it often results in correction clicks being distributed across various regions or lesions, limiting the ability to perform precise lesion-wise refinement — particularly problematic in high tumor burden WB-MRI scans.

To address this limitation, we implemented a lesion-wise evaluation pipeline (Algorithm \ref{alg:evaluation}), specifically tailored for multi-lesion scenarios. Our pipeline relied on connected component analysis to extract individual lesions from the ground truth multi-lesion mask. A subset of the largest lesions was selected for refinement of the segmentation accuracy by interaction. Each lesion of the subset was refined independently by iteratively placing simulated user clicks at regions of maximum discrepancy between the prediction and the ground truth. Due to potential concave or complex lesion shapes, the initial click location (e.g., geometric center of the error region) may fall outside the actual segmentation error. In such cases, the click was projected to the nearest voxel within the error area to ensure effective interaction. The final multi-lesion semantic segmentation mask was obtained by merging the refined single-lesion instance segmentation masks.  To reduce noise and improve structural coherence, small segmentation artefacts were removed by filtering connected components with a volume smaller than a predefined threshold ($V_{\text{thresh}}$).

Key parameters of this pipeline included the number of largest lesions selected ($L_{\text{chosen}}$), the number of simulated interactions per lesion ($C_{\text{chosen}}$), the artefact removal volume threshold ($V_{\text{thresh}}$), and the set of evaluation metrics ($S_{\text{metrics}}$).

% Algorithm start
\begin{figure}[ht]
\scriptsize % or \footnotesize or \scriptsize
\begin{algorithm}[H]
\label{alg:evaluation}
\caption{Lesion-wise evaluation pipeline for assessment of interactive segmentation models}
\KwIn{Multi-lesion semantic ground truth mask $M_{\text{GT}}$, number of largest lesions to be selected $L_{\text{chosen}}$, number of simulated interactions per lesion $C_{\text{chosen}}$, artefact removal volume threshold $V_{\text{thresh}}$}
\KwOut{Set of calculated metrics $S_{\text{metrics}}$}

Perform connected component analysis on $M_{\text{GT}}$ to extract the instance segmentation mask\;
Select $L_{\text{chosen}}$ largest lesions from the instance segmentation mask\;
Initialize empty multi-lesion semantic prediction mask $M_{\text{pred}}$\;

\ForEach{lesion $l$ in $L_{\text{chosen}}$}{
  Get single-lesion instance ground truth mask $m_{\text{GT}}^{(l)}$ for lesion $l$\;
  Initialize empty single-lesion instance prediction mask $\hat{m}^{(l)}$\;
  
  \For{$i \gets 1$ \KwTo $C_{\text{chosen}}$}{
    Compare $\hat{m}^{(l)}$ and $m_{\text{GT}}^{(l)}$ to find largest error region\;
    Simulate a click at this region (positive/negative depending on false negative/positive)\;
    Apply model with simulated clicks to obtain an updated prediction mask $\hat{m}^{(l)}$\;
  }
  
  Insert $\hat{m}^{(l)}$ into $M_{\text{pred}}$\;
}

Remove small connected components $< V_{\text{thresh}}$ from $M_{\text{pred}}$\;
Compute evaluation metrics $S_{\text{metrics}}$ between $M_{\text{pred}}$ and $M_{\text{GT}}$\;
\end{algorithm}
\end{figure}
% Algorithm end

\subsubsection{Metrics and Statistical Analysis}
\label{subsubsec:metrics_and_statistical_analysis}

The primary metric used was the scan-wise Dice Similarity Coefficient (DSC), which measures the overlap between the predicted multi-lesion semantic segmentation mask and the ground truth mask over an entire WB-MRI scan. Scan-wise DSC was computed per patient. 

To assess the ability of each model to correctly detect individual lesions, we computed the lesion detection rate using the F1 score. A lesion in the ground truth was considered successfully detected if any predicted lesion overlapped with it above a predefined IoU threshold, set to 0.1. Individual lesions were identified via connected component analysis in both predicted and ground truth masks. 

The choice of an IoU threshold of 0.1 reflects a trade-off between sensitivity and specificity in lesion detection. This low threshold ensures that even partial overlaps with the ground truth are recognized as detections, avoiding the brittleness of a single-voxel criterion while remaining more robust than specifying a fixed number of overlapping pixels. 

For each correctly detected lesion, we computed the lesion-wise DSC at the connected component level. The average lesion-wise DSC was calculated across all correctly detected lesions per scan. Lesions with no corresponding prediction were excluded from this metric to avoid skewing the average with zeros, as our goal was to evaluate segmentation quality rather than sensitivity (already reflected by the F1 score).

To suppress noise and ensure clinical relevance, connected components with a physical volume $V_{\text{thresh}}$ below 1~cm\textsuperscript{3} were excluded from both prediction and ground truth masks. Prior studies varied substantially in their thresholding strategies: some excluded tumors smaller than 75~cm\textsuperscript{3}~\cite{solomon_automated_2004, weizman_interactive_2012}, while others included lesions down to 5~cm\textsuperscript{3}~\cite{weizman_pnist_2014, wu_deep_2020}. In our study, we aimed to filter out implausibly small objects that were unlikely to represent NFs, while retaining sensitivity to the smallest true-positive lesions, particularly in Test Set~3, which contained low tumor burden cases. 

All models and their variants were trained three times using three-fold cross-validation exclusively on the training set. Different random seeds were used for each fold to ensure training diversity. This approach was adopted to ensure that model performance estimates were robust to variability in training data splits.

Performance metrics were computed on the validation fold and reported as mean ± standard deviation across the three folds. For test set evaluations, each of the three trained models was applied independently to all test sets, and the resulting predictions were evaluated separately and aggregated to yield mean ± standard deviation. To assess the statistical significance of performance differences between models, we employed the Wilcoxon signed-rank test for paired comparisons of metrics. When comparing MOIS-SAM2 to multiple baselines, the Bonferroni correction was applied to account for multiple comparisons, with a corrected significance threshold of $p < 0.01$.

\subsubsection{Experiments}
\label{subsubsec:experiments}

We carried out experiments to evaluate the proposed MOIS-SAM2 model across: architecture ablation, exemplars count efficiency, baseline benchmarking, and SAM2-vs-MOIS-SAM2 interaction efficiency.

\paragraph{Experiment 1: Architecture ablation}

This experiment assessed how architecture components of MOIS-SAM2 contribute to NF segmentation performance. The following model variants were evaluated:

 \begin{itemize}
     \item Approach 1: SAM2 (multi-lesion segmentation training only). Original SAM2 architecture without exemplars. The model was trained to predict a multi-lesion semantic segmentation mask using click prompts. Single-lesion instance segmentation was not supported. This option tested whether the unmodified SAM2 can propagate a single click to multiple lesions. 
     \item Approach 2: SAM2 with exemplar bank, no exemplar attention, full exemplar structure. The model was trained to perform both single-lesion instance segmentation from click prompts and multi-lesion semantic segmentation from exemplars. The exemplars were passed to the memory attention. This option tested whether memory attention can handle both memory-based scan-wise propagation and exemplar-based semantic propagation at the same time.
     \item Approach 3: MOIS-SAM2, exemplar without object pointer encoding lesion identity. This option evaluated whether exemplar-based semantic propagation is compromised without explicit lesion identifiers.
     \item Approach 4: MOIS-SAM2, exemplar without object slice index. This option tested whether including inter-slice location affects the semantic propagation of exemplars.
     \item Approach 5: MOIS-SAM2 (proposed). Complete MOIS-SAM2 with exemplar attention, exemplar bank, and full exemplar structure.
 \end{itemize}

Performance was evaluated using the lesion-wise evaluation pipeline with $L_{\text{chosen}}$ = 20 prompted lesions and $C_{\text{chosen}}$ = 3 simulated user clicks per lesion. The choice of $L_{\text{chosen}} = 20$ was inspired by the previous study on interactive segmentation of NF with DINs \cite{zhang_dins_2022}. Given that NF1 patients may present with hundreds of lesions, we opted to distribute the interaction budget across 20 distinct lesions. For each lesion, three simulated user clicks were used: one initial click followed by two corrective clicks, reflecting a realistic interactive refinement process.

\paragraph{Experiment 2: Effect of exemplars count}

This experiment evaluated how segmentation accuracy is influenced by the number of prompted exemplars. We tested the MOIS-SAM2 model on the same validation folds as in Experiment 1.

Performance was assessed using the lesion-wise evaluation pipeline with varying the number of prompted lesions $L_{\text{chosen}}$ from 1 to 10. We fixed the number of simulated user clicks per lesion $C_{\text{chosen}}$ to 1 to reflect a minimal interaction scenario, to isolate the impact of exemplar quantity, and to prevent confounding from excessive corrections. We reported metrics scan-wise DSC, lesion detection F1 score, and lesion-wise DSC as a function of the number of prompted lesions.

\paragraph{Experiment 3: Benchmarking against baseline models}

We benchmarked MOIS-SAM2 against a range of models, both interactive and automated:

 \begin{itemize}
     \item U-Net and nnU-Net: convolution-based automated models trained from scratch.
     \item DINs and SW-FastEdit: convolution-based interactive models fine-tuned from publicly available checkpoints.
     \item SAM2 and VISTA3D: transformer-based interactive models fine-tuned from publicly available checkpoints.
 \end{itemize}

All models were evaluated on the four distinct test sets, enabling analysis under both in-domain and out-of-distribution conditions. Each model was tested using $L_{\text{chosen}}$ = 20 prompted lesions and $C_{\text{chosen}}$ = 3 simulated clicks per lesion.

To analyze generalization under domain shift for each model, we compared scan-wise DSC distributions between Test Set 1 (in-domain) and each of the other Test Sets (2–4) using the Mann–Whitney U test.

\paragraph{Experiment 4: Interaction efficiency}

This experiment evaluated the interaction efficiency of SAM2 and MOIS-SAM2. For each model, we varied the number of prompted lesions $L_{\text{chosen}}$ from 1 to 10, using a fixed number of simulated user clicks per lesion ($C_{\text{chosen}} = 1$). This setup reflected a lightweight interaction workflow in which the user provides a single prompt per lesion.

\subsubsection{Clinical Validity}
\label{subsubsec:clinical_validity}

To assess the clinical relevance of the proposed segmentation pipeline and compare its performance to human experts, we conducted a small inter-reader variability analysis.

NFs in four representative NF1 patients were independently manually annotated with no time constraints. The annotators represented a spectrum of clinical expertise:

\begin{itemize}
    \item Three radiologists with experience in NF imaging (I.R., M.-L.S., L.W.)
    \item One fourth year resident in radiology without NF-specific expertise (S.G.)
    \item One medical student without radiological training (Stud.)
\end{itemize}

For comparison with the model, a non-medical user (G.K.) provided $C_{\text{chosen}}$ = 1 user click per each of the $L_{\text{chosen}}$ = 20 largest lesions. This setup was chosen to reflect a practical, low-effort interaction scenario and to maintain consistency with the main evaluation protocol. These prompts were used to guide MOIS-SAM2 to perform multi-lesion segmentation via exemplar-based semantic propagation.

The dataset included:
\begin{itemize}
    \item Two in-domain cases from Test Set~1:
    \begin{itemize}
        \item A 28-year-old male (tumor burden $\sim$1000~cm$^3$)
        \item A 61-year-old female ($\sim$700~cm$^3$)
    \end{itemize}
    \item Two out-of-distribution cases not present in training or test sets:
    \begin{itemize}
        \item A 36-year-old female ($\sim$200~cm$^3$)
        \item A 16-year-old male ($\sim$150~cm$^3$)
    \end{itemize}
\end{itemize}

For each case, the multi-lesion segmentation masks from each human annotator and the MOIS-SAM2 model were compared pairwise using the scan-wise DSC.

\subsection{Implementation Details}
\label{subsec:implementation_details}

\subsubsection{Training Parameters}
\label{subsubsec:training_parameters}

The training hyperparameters of the proposed MOIS-SAM2 model largely followed the default configuration of SAM2. 

The dedicated exemplar attention module consisted of four transformer layers with rotary positional encoding (RoPE), each using 256-dimensional embeddings and a single attention head. This configuration followed the design of the memory attention module in SAM2. The exemplar bank was configured to store up to 10 exemplars, a number chosen based on GPU memory constraints. Each exemplar included both the object pointer and object slice index to preserve lesion identity and spatial localization.

All backbone modules (image encoder, prompt encoder, memory encoder, memory attention, and mask decoder) were initialized from a pretrained SAM2 checkpoint and fine-tuned during training. The newly introduced exemplar attention module was initialized using weights copied from the SAM2 memory attention module. This strategy mitigated cold-start effects associated with the newly introduced module.

Training samples consisted of 4 sequential coronal slices with up to 3 lesions prompted per sample. These values were selected to ensure compatibility with available GPU memory and feasible training times. For each prompted lesion, 1 initial and 6 corrective user clicks were simulated, consistent with the SAM2 training strategy. Data augmentation included random affine transformations (rotation $\pm10^\circ$, shear $\pm5^\circ$), horizontal flipping, random cropping, and resizing to $1024 \times 1024$ pixels.

The model was trained for 100 epochs using the AdamW optimizer with a learning rate of $5 \times 10^{-6}$ and a batch size of 1. Training convergence was monitored using the mean validation scan-wise DSC, computed on the held-out fold during 3-fold cross-validation. No early stopping was used. The model was optimized using a composite loss function (Section~\ref{subsec:training_strategy}).

All training and evaluation scripts, along with full hyperparameter and architectural specifications, are available in the public GitHub repository: \href{https://github.com/IPMI-ICNS-UKE/MOIS_SAM2_NF}{GitHub}.

\subsubsection{Software and Hardware}
\label{subsubsec:software_and_hardware}

The training pipeline was implemented in Python 3.10 using PyTorch. The lesion-wise evaluation pipeline utilized MONAI, NumPy, SciPy, and SimpleITK, with parallel lesion-wise processing implemented via joblib. Baseline interactive segmentation models were integrated into the evaluation pipeline following the MONAI framework logic to ensure consistent and reproducible benchmarking. The MOIS-SAM2 interactive segmentation pipeline was further integrated into MONAI Label \cite{diaz-pinto_monai_2024}, enabling its usability via interactive annotation platforms such as 3D Slicer \cite{fedorov_3d_2012}.

All experiments were conducted on a workstation running Ubuntu 22.04.5 LTS, equipped with an AMD Ryzen Threadripper Pro 3975WX CPU and two NVIDIA RTX A6000 GPUs.

\section{Results}
\label{sec:results}

\subsection{Architecture Ablation}
\label{subsec:architecture_ablation_results}

The contribution of individual architecture components of MOIS-SAM2 to NF segmentation performance is summarized in Table \ref{tab:architecture_ablation}. Approach 1, corresponding to the original SAM2 architecture trained solely for multi-lesion semantic segmentation without exemplar-based semantic propagation, showed a scan-wise DSC of 0.50 ± 0.16. 

Approach 2, the original SAM2 architecture which reused the memory attention module for both memory-based scan-wise propagation and exemplar-based semantic propagation, failed to leverage exemplar guidance effectively (scan-wise DSC of 0.49 ± 0.18), resulting in no improvement over Approach 1. This suggests that a shared memory attention module is insufficient for exemplar-based semantic propagation.

Approaches 3 and 4 evaluated the effect of removing components from the exemplar structure. Removing the object pointer (Approach 3) led to a degradation across all metrics (scan-wise DSC of 0.47 ± 0.18), indicating that explicit lesion identity is critical for accurate exemplar-based semantic propagation. In contrast, excluding the object slice index (Approach 4) had little impact on the metrics compared to the full exemplar structure in Approach 5 (scan-wise DSC of 0.57 ± 0.16).

Approach 5, the full MOIS-SAM2 model with dedicated exemplar attention, an exemplar bank, and the full exemplar structure, achieved the best scan-wise DSC (0.59 ± 0.13) and the highest lesion detection F1 score (0.73 ± 0.10), while maintaining strong lesion-wise DSC (0.69 ± 0.11). These results support the conclusion that dedicated exemplar mechanisms significantly enhance multi-lesion semantic segmentation performance without compromising lesion-wise accuracy.

\begin{table}[t]
\centering
\scriptsize
\begin{tabular}{>{\raggedright\arraybackslash}p{0.16\linewidth}>{\centering\arraybackslash}p{0.13\linewidth}>{\centering\arraybackslash}p{0.13\linewidth}>{\centering\arraybackslash}p{0.13\linewidth}>{\centering\arraybackslash}p{0.15\linewidth}>{\centering\arraybackslash}p{0.13\linewidth}}
\textbf{Metric (Mean~$\pm$~SD)} & \textbf{Approach 1} & \textbf{Approach 2} & \textbf{Approach 3} & \textbf{Approach 4} & \textbf{Approach 5 (Proposed)}\\
\hline
Scan-wise DSC& 0.50 $\pm$ 0.16\textsuperscript{*} & 0.49 $\pm$ 0.18\textsuperscript{*} & 0.47 $\pm$ 0.18\textsuperscript{*} & 0.57 $\pm$ 0.16 & \textbf{0.59 $\pm$ 0.13} \\

Lesion F1 score  & 0.70 $\pm$ 0.12 & 0.72 $\pm$ 0.11 & 0.65 $\pm$ 0.14\textsuperscript{*} & 0.72 $\pm$ 0.12 & \textbf{0.73 $\pm$ 0.10} \\

Lesion-wise DSC & 0.64 $\pm$ 0.20 & 0.64 $\pm$ 0.15 & 0.62 $\pm$ 0.23 & \textbf{0.69 $\pm$ 0.12} & \textbf{0.69 $\pm$ 0.11} \\

\hline
\multicolumn{6}{p{0.95\linewidth}}{\textsuperscript{*}Statistically significant difference from Approach 5 (Wilcoxon signed-rank test with Bonferroni correction). DSC – Dice Similarity Coefficient; SD – standard deviation.}\\  
\end{tabular}
\vspace{1ex}
\caption{ 
Segmentation performance of model variants in the architecture ablation study. (1) SAM2 without exemplar mechanism (trained for multi-lesion segmentation only); (2) SAM2 with exemplars but no exemplar attention (exemplars routed through memory attention); (3) MOIS-SAM2 without object pointer encoding lesion identity; (4) MOIS-SAM2 without object slice index; (5) Full MOIS-SAM2 with exemplar attention, exemplar bank, and a full exemplar structure (proposed). Evaluation used 20 prompted lesions and 3 simulated user clicks per lesion.
}
\label{tab:architecture_ablation}
\end{table}

\subsection{Effect of Exemplars Count}
\label{subsec:effect_exemplars_count_results}

Analysis of segmentation performance as a function of the number of prompted exemplars showed improvement across all metrics when increasing from 1 to 2 exemplars. Scan-wise DSC increased by 7.9\% (from 0.51 to 0.57), lesion detection F1 score by 13.7\% (from 0.51 to 0.58), and lesion-wise DSC by 13.5\% (from 0.37 to 0.42), highlighting the benefit of exemplar-based semantic propagation even at a low exemplars count. Notably, the metrics exhibited the highest variability with a single prompted exemplar, likely due to high sensitivity to the initial prompt location. Performance steadily improved with additional exemplars and reached a plateau after 5–6 prompts across all metrics: 0.59 of scan-wise DSC, 0.64 of lesion detection F1 score, 0.50 of lesion-wise DSC. This saturation effect indicates that MOIS-SAM2 can effectively perform semantic propagation to unprompted lesions after prompting only a small number of exemplar lesions.

\subsection{Benchmarking against Baseline Models}
\label{subsec:benchmarking_results}

\subsubsection{Quantitative Results}

Benchmarking the proposed MOIS-SAM2 model against six baseline and state-of-the-art segmentation models across all test sets (Table \ref{tab:benchmarking_baselines}) showed that our approach achieved the best overall segmentation performance in terms of scan-wise DSC and lesion detection F1 score.  

On Test Set 1 (in-domain), MOIS-SAM2 (Model 7) achieved a scan-wise DSC of 0.60 ± 0.17 and a lesion detection F1 score of 0.74 ± 0.17, outperforming the baseline automated model nnU-Net (Model 2) by 11.1\% of scan-wise DSC (0.54 ± 0.24) and 29.8\% of lesion detection F1 score (0.57 ± 0.23). It also exceeded the state-of-the-art interactive model VISTA3D (Model 6) by 36.4\% (0.44 ± 0.09) and 27.6\% (0.58 ± 0.16), respectively. 

However, the proposed model showed a lower lesion-wise DSC than the automated models (Models 1 and 2). This apparent discrepancy arises from the complementary behavior of the lesion detection F1 score and lesion-wise DSC. Automated models tend to oversegment lesions (Figure \ref{fig:Figure_3} - (a) Model 1), leading to higher false positive rates and lower F1 scores. However, this oversegmentation increases the likelihood of overlap with true lesion regions, which inflates the lesion-wise DSC (computed only on correctly detected lesions). In contrast, MOIS-SAM2 detects substantially more lesions with greater precision, resulting in a more conservative segmentation that may slightly lower lesion-wise DSC while achieving superior overall performance.

On Test Set 2 (MRI field strength variation), MOIS-SAM2 (Model 7) outperformed all baselines in scan-wise DSC (0.53 ± 0.13) and lesion detection F1 score (0.78 ± 0.13). MOIS-SAM2 achieved a 12.8\% improvement in scan-wise DSC and 30.0\% in F1 score compared to nnU-Net (Model 2), which had metric values of 0.47 ± 0.16 and 0.60 ± 0.20, respectively. The gains against VISTA3D (Model 6) were even larger: 43.2\% (0.37 ± 0.11) in scan-wise DSC and 44.4\% (0.54 ± 0.13) in F1 score. As on Test Set 1, the automated models – U-Net (Model 1, 0.67 ± 0.29) and nnU-Net (Model 2, 0.66 ± 0.35) – achieved higher lesion-wise DSC than MOIS-SAM2 (0.58 ± 0.36).

On Test Set 3 (patients with low tumor burden), MOIS-SAM2 (Model 7) demonstrated exceptional robustness in sparse lesion scenarios. It achieved a scan-wise DSC of 0.61 ± 0.40 and a lesion detection F1 score of 0.41 ± 0.23, substantially outperforming both automated and interactive segmentation models. Despite this, the lesion-wise DSC for MOIS-SAM2 (0.72 ± 0.26) was on par with U-Net (0.70 ± 0.22) and nnU-Net (0.69 ± 0.34), with no statistically significant differences. These findings underscore that MOIS-SAM2 retains strong per-lesion segmentation accuracy while detecting significantly more lesions – particularly important in low tumor burden cases, where subtle lesions may be easily missed. 

Test Set 4 represents a severe domain shift due to a different clinical site and scanner vendor. The domain gap is clearly visible in the images themselves (Figure \ref{fig:Figure_3} - case 4), with differing contrast and noise characteristics due to acquisition on a Philips scanner, which was not present in the training data. Despite this, MOIS-SAM2 (Model 7) remained robust, owing to its exemplar-based semantic propagation mechanism. Since exemplars were added from the test scan itself during interaction, the model could adapt to features from this unseen domain. While its absolute performance was lower than on Test Sets 1-3, MOIS-SAM2 (Model 7) still achieved the best performance with a scan-wise DSC of 0.50 ± 0.19 and a lesion detection F1 score of 0.62 ± 0.15. Compared to the best-performing automated baseline, nnU-Net (Model 2), this corresponds to a 25.0\% improvement in scan-wise DSC (0.40 ± 0.21) and a 21.6\% improvement in F1 score (0.51 ± 0.17). Against the interactive transformer-based baseline SAM2 (Model 5), MOIS-SAM2 (Model 7) showed a 35.1\% increase in scan-wise DSC (0.37 ± 0.22) and a 14.8\% increase in F1 score (0.54 ± 0.28). 

\begin{table}
\centering
\footnotesize
\renewcommand{\arraystretch}{1.0}
\setlength{\tabcolsep}{3pt}
\begin{tabular}{
>{\raggedright\arraybackslash}p{0.18\linewidth}
>{\centering\arraybackslash}p{0.1\linewidth}
>{\centering\arraybackslash}p{0.1\linewidth}
>{\centering\arraybackslash}p{0.1\linewidth}
>{\centering\arraybackslash}p{0.1\linewidth}
>{\centering\arraybackslash}p{0.1\linewidth}
>{\centering\arraybackslash}p{0.1\linewidth}
>{\centering\arraybackslash}p{0.1\linewidth}}
\hline
\textbf{Metric (Mean$\pm$SD)} & \textbf{Model 1} & \textbf{Model 2} & \textbf{Model 3} & \textbf{Model 4} & \textbf{Model 5} & \textbf{Model 6} & \textbf{Model 7} \\

\hline
\multicolumn{8}{p{0.95\linewidth}}{\textbf{Test Set 1 – In-domain}} \\
\hline
Scan-wise DSC & 0.39 ± 0.23\textsuperscript{*}& 0.54 ± 0.24\textsuperscript{*}& 0.37 ± 0.17\textsuperscript{*}& 0.34 ± 0.12\textsuperscript{*}& 0.35 ± 0.19\textsuperscript{*}& 0.44 ± 0.09\textsuperscript{*}& \textbf{0.60 ± 0.17} \\
Lesion F1 score & 0.38 ± 0.26\textsuperscript{*}& 0.57 ± 0.23\textsuperscript{*}& 0.51 ± 0.15\textsuperscript{*}& 0.55 ± 0.20\textsuperscript{*}& 0.51 ± 0.34\textsuperscript{*}& 0.58 ± 0.16\textsuperscript{*}& \textbf{0.74 ± 0.17} \\
Lesion-wise DSC & 0.57 ± 0.34& \textbf{0.62 ± 0.38\textsuperscript{*}}& 0.17 ± 0.28\textsuperscript{*}& 0.20 ± 0.31\textsuperscript{*}& 0.07 ± 0.21\textsuperscript{*}& 0.23 ± 0.34\textsuperscript{*}& 0.55 ± 0.36 \\

\hline
\multicolumn{8}{p{0.95\linewidth}}{\textbf{Test Set 2 – MRI Field Strength Variation}} \\
\hline
Scan-wise DSC & 0.21 ± 0.14\textsuperscript{*}& 0.47 ± 0.16\textsuperscript{*}& 0.44 ± 0.15\textsuperscript{*}& 0.31 ± 0.14\textsuperscript{*}& 0.33 ± 0.24\textsuperscript{*}& 0.37 ± 0.11\textsuperscript{*}& \textbf{0.53 ± 0.13} \\
Lesion F1 score & 0.36 ± 0.24\textsuperscript{*}& 0.60 ± 0.20\textsuperscript{*}& 0.66 ± 0.18\textsuperscript{*}& 0.49 ± 0.20\textsuperscript{*}& 0.61 ± 0.31\textsuperscript{*}& 0.54 ± 0.13\textsuperscript{*}& \textbf{0.78 ± 0.13} \\
Lesion-wise DSC & \textbf{0.67 ± 0.29\textsuperscript{*}}& 0.66 ± 0.35\textsuperscript{*}& 0.32 ± 0.35\textsuperscript{*}& 0.23 ± 0.34\textsuperscript{*}& 0.15 ± 0.30\textsuperscript{*}& 0.30 ± 0.37\textsuperscript{*}& 0.58 ± 0.36 \\

\hline
\multicolumn{8}{p{0.95\linewidth}}{\textbf{Test Set 3 – Low Tumor Burden}} \\
\hline
Scan-wise DSC & 0.01 ± 0.02\textsuperscript{*}& 0.15 ± 0.26\textsuperscript{*}& 0.09 ± 0.16\textsuperscript{*}& 0.21 ± 0.34\textsuperscript{*}& 0.54 ± 0.43& 0.31 ± 0.35\textsuperscript{*}& \textbf{0.61 ± 0.40} \\
Lesion F1 score & 0.02 ± 0.02\textsuperscript{*}& 0.16 ± 0.28\textsuperscript{*}& 0.18 ± 0.23\textsuperscript{*}& 0.16 ± 0.28\textsuperscript{*}& 0.35 ± 0.46& 0.23 ± 0.31\textsuperscript{*}& \textbf{0.41 ± 0.23} \\
Lesion-wise DSC & 0.70 ± 0.22& 0.69 ± 0.34& 0.63 ± 0.25\textsuperscript{*}& 0.54 ± 0.37\textsuperscript{*}& 0.52 ± 0.40\textsuperscript{*}& 0.68 ± 0.29& \textbf{0.72 ± 0.26} \\

\hline
\multicolumn{8}{p{0.95\linewidth}}{\textbf{Test Set 4 – Clinical Site and Scanner Vendor Variation}} \\
\hline
Scan-wise DSC & 0.23 ± 0.20\textsuperscript{*}& 0.40 ± 0.21\textsuperscript{*}& 0.34 ± 0.15\textsuperscript{*}& 0.21 ± 0.13\textsuperscript{*}& 0.37 ± 0.22\textsuperscript{*}& 0.23 ± 0.12\textsuperscript{*}& \textbf{0.50 ± 0.19} \\
Lesion F1 score & 0.30 ± 0.20\textsuperscript{*}& 0.51 ± 0.17\textsuperscript{*}& 0.56 ± 0.20& 0.46 ± 0.21\textsuperscript{*}& 0.54 ± 0.28& 0.53 ± 0.19& \textbf{0.62 ± 0.15} \\
Lesion-wise DSC & 0.61 ± 0.32& 0.46 ± 0.34\textsuperscript{*}& 0.15 ± 0.26\textsuperscript{*}& 0.14 ± 0.26\textsuperscript{*}& 0.14 ± 0.29\textsuperscript{*}& 0.16 ± 0.27\textsuperscript{*}& \textbf{0.62 ± 0.34} \\
\hline

\multicolumn{8}{p{0.95\linewidth}}{\textsuperscript{*}Statistically significant difference from the proposed Model 7 (Wilcoxon signed-rank test with Bonferroni correction). DSC – Dice Similarity Coefficient; SD – standard deviation.}\\  
\end{tabular}
\vspace{1ex}
\caption{Segmentation performance of MOIS-SAM2 and six baseline models on the four test sets. Automated convolution-based models: (1) U-Net, (2) nnU-Net; interactive convolution-based models: (3) DINs, (4) SW-FastEdit; interactive transformer-based models: (5) SAM2, (6) VISTA3D; and (7) proposed MOIS-SAM2. Evaluation used 20 prompted lesions and 3 simulated user clicks per lesion.
}
\label{tab:benchmarking_baselines}
\end{table}

\subsubsection{Qualitative Results}

\begin{figure}
     \centering
     \includegraphics[width=0.6\linewidth]{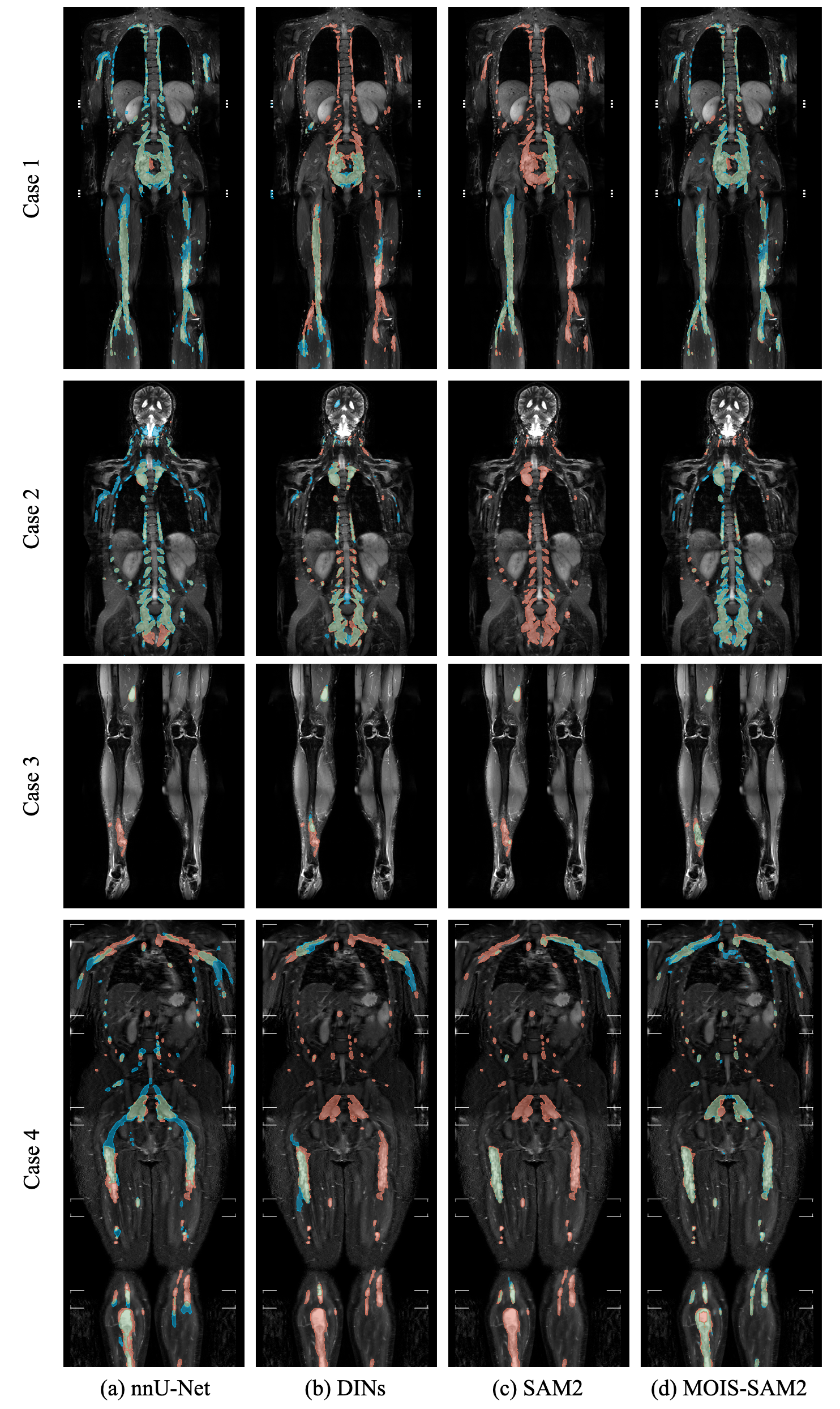}
     \caption{Qualitative comparison of segmentation results from MOIS-SAM2 and three baseline models across four representative test cases. Each row corresponds to a single patient case, and each column to a respective model: (a) nnU-Net (automated convolution-based), (b) DINs (interactive convolution-based), (c) SAM2 (interactive transformer-based), and (d) MOIS-SAM2 (proposed). Cases: (1) in-domain, (2) MRI field strength variation, (3) low tumor burden, (4) clinical site and scanner vendor variation. True positives are shown in green, false positives in blue, and false negatives in red. Evaluation used 20 prompted lesions and 3 simulated user clicks per lesion.}
     \label{fig:Figure_3}
 \end{figure}

To qualitatively assess segmentation results, we compared four representative models – nnU-Net (automated convolution-based model), DINs (interactive convolution-based model), SAM2 (interactive transformer-based model), and the proposed MOIS-SAM2 – on four exemplar patients (Figure \ref{fig:Figure_3}), each corresponding to one of the Test Sets 1-4.

On in-domain case 1, nnU-Net displayed strong lesion coverage but substantial false positives in paraspinal and lower extremity musculature, and incomplete coverage of the central pelvic plexiform lesion. DINs undersegmented spinal and femoral lesions, missing axillary tumors entirely. SAM2 produced clean but highly localized segmentations, capturing some lesions while missing many unprompted ones in the spine, pelvis, and shoulders. MOIS-SAM2 demonstrated the most balanced segmentation, accurately identifying thoracolumbar, pelvic, and axillary lesions with fewer false positives. However, some oversegmentation was still observed in the proximal thigh and gluteal muscles, and small hyperintense structures near the kidneys were occasionally mislabeled as lesions.

On case 2 with MRI field strength variation, nnU-Net maintained good coverage of paraspinal lesions but oversegmented surrounding musculature, with frequent false positives in the cervical and axillary regions. DINs struggled with hyperintense brain structure, produced incomplete and inconsistent segmentations across spine and pelvis, along with false positives in the abdomen. SAM2 failed to propagate segmentation beyond prompted areas, missing the majority of thoracolumbar and pelvic lesions. MOIS-SAM2 more reliably captured spinal, pelvic, and retroperitoneal lesions. However, several limitations were evident: undersegmentation was present in the posterior neck and shoulder girdle.

On low tumor burden case 3, nnU-Net accurately detected one lesion but missed the other two entirely and produced false positives in the upper thigh. DINs correctly identified two lesions but undersegmented one of them and failed to detect the third. SAM2 achieved high local accuracy on one lesion, partially segmented the second, and completely missed the third. MOIS-SAM2 successfully identified two lesions – though it missed the peripheral region of one of them and failed to detect the third.

On case 4 with a clinical site and scanner vendor variation, nnU-Net showed extensive false positives, missed lesions, and provided inconsistent masks across gluteal, thigh, and chest regions. DINs displayed coarse segmentation, with high false negative rates. SAM2 captured only a few prompted lesions and missed the majority of the tumor burden. MOIS-SAM2  segmented spinal, pelvic, thigh, and axillary lesions with tight boundaries, but also experienced oversegmentation in shoulder and had false negative areas in pelvis and one of the legs.

While nnU-Net and DINs were prone to false positives and poor generalization, SAM2 struggled to scale beyond prompted lesions, MOIS-SAM2 delivered the best visual balance between lesion recall and precision.

\subsubsection{Domain Generalization}

\begin{figure}[t]
     \centering
     \includegraphics[width=0.85\linewidth]{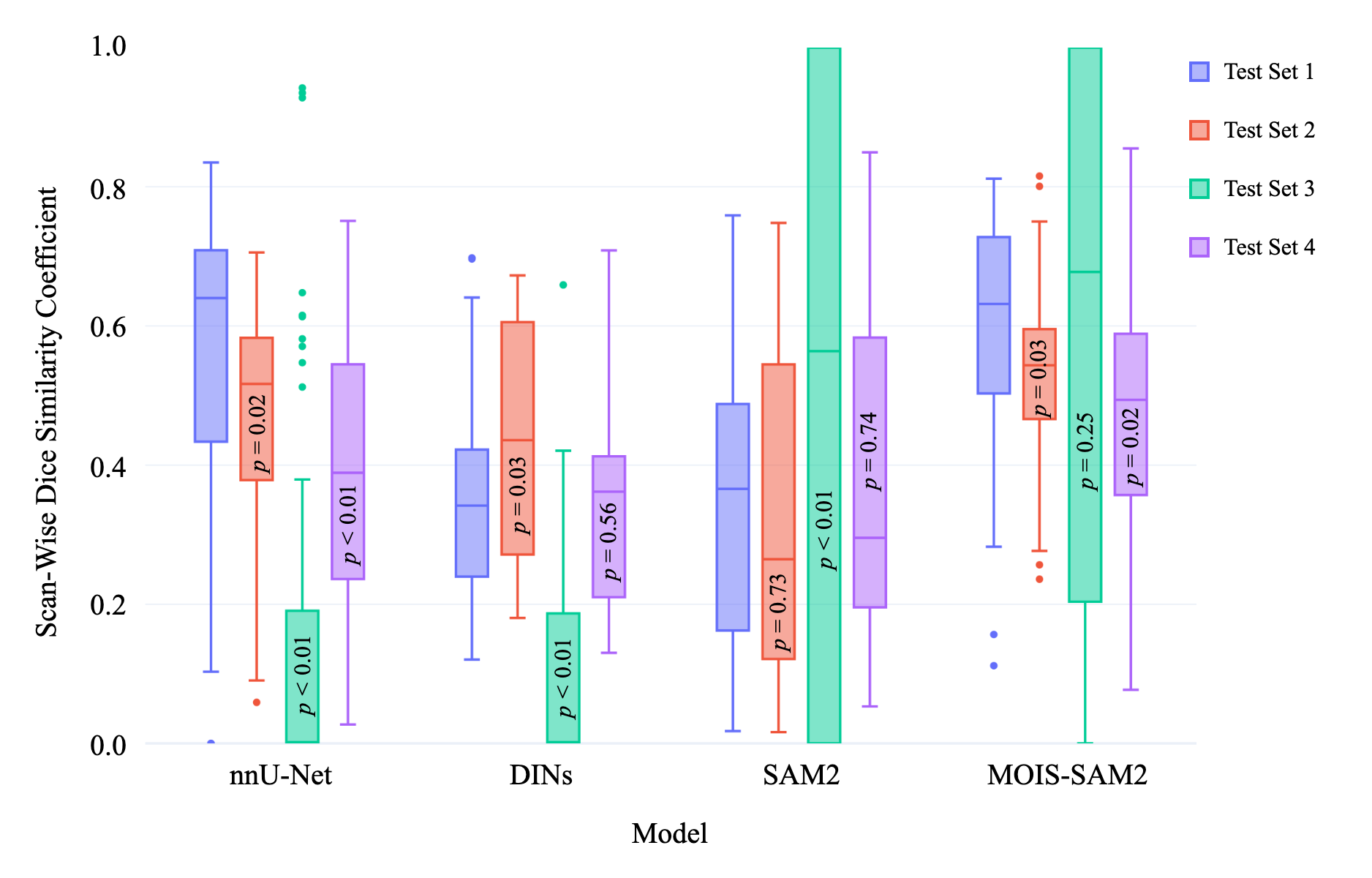}
     \caption{Domain shift analysis of MOIS-SAM2 and three baseline models. Scan-wise Dice Similarity Coefficient (DSC) distributions are shown for each model (nnU-Net, DINs, SAM2, MOIS-SAM2) across four test sets: Test Set 1 (in-domain), Test Set 2 (MRI field strength variation), Test Set 3 (low tumor burden), and Test Set 4 (clinical site and scanner vendor variation). Boxes indicate the interquartile range of scan-wise DSC with medians. P-values from Mann–Whitney U tests compare each out-of-distribution Test Set 2-4 to Test Set 1.}
     \label{fig:Figure_4}
 \end{figure}

The analysis of model robustness to domain shift (Figure \ref{fig:Figure_4}) showed that nnU-Net exhibited significant performance degradation on Test Sets 3 and 4 (p < 0.01 for both) and no significant change on Test Set 2 (p = 0.02). This indicates low robustness of nnU-Net in handling low tumor burden cases and high sensitivity to variations in clinical site and scanner vendor. DINs also suffered a significant drop on Test Set 3 (p < 0.01), but showed no significant performance changes on Test Sets 2 and 4 (p = 0.03 and 0.56, respectively), reflecting relatively stable behaviour, potentially due to the use of user interactions. SAM2 showed a notable improvement on Test Set 3 (p < 0.01), likely because of its prompting mechanism. It exhibited no significant shift on other test sets, indicating stable generalization across domains. MOIS-SAM2 did not show a significant drop on any of the test sets (p > 0.02 in all comparisons). This demonstrates more robust generalization to domain shift than in compared models.

\subsection{Interaction Efficiency}
\label{subsec:interaction_efficiency}

\begin{figure}[t]
     \centering
     \includegraphics[width=0.60\linewidth]{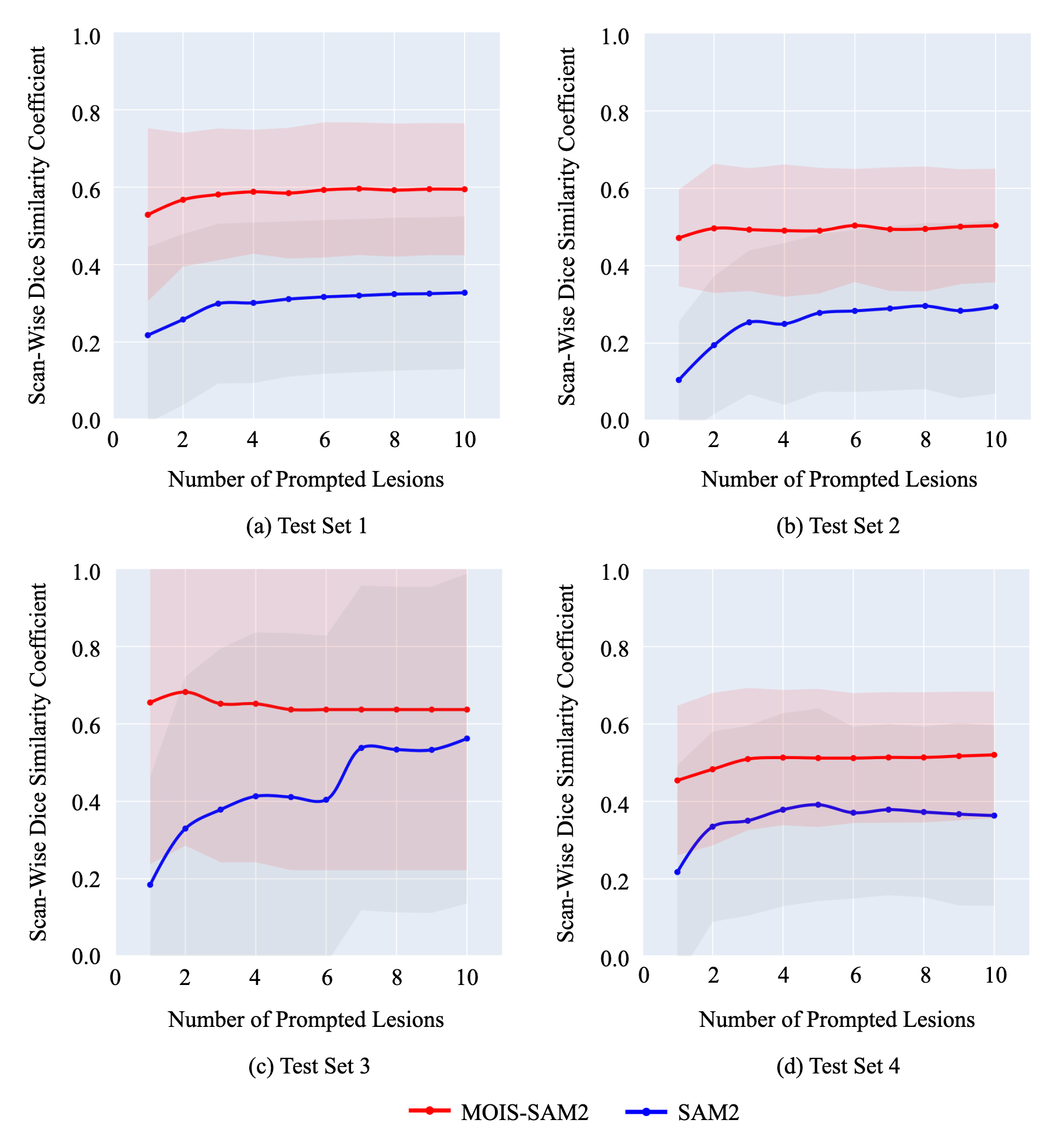}
     \caption{Interaction efficiency of SAM2 vs MOIS-SAM2 across test sets. Scan-wise Dice Similarity Coefficient (DSC) as a function of the number of prompted lesions for SAM2 (blue) and MOIS-SAM2 (red), evaluated on four test sets: (a) Test Set 1 – in-domain, (b) Test Set 2 – MRI field strength variation, (c) Test Set 3 – low tumor burden, and (d) Test Set 4 – clinical site and scanner vendor variation. Evaluation used from 1 to 10 prompted lesions and 1 simulated user click per lesion. Results are shown with standard deviation bands.}
     \label{fig:Figure_5}
 \end{figure}

The comparison of interaction efficiency between SAM2 and MOIS-SAM2 (Figure \ref{fig:Figure_5}) showed that MOIS-SAM2 consistently outperformed SAM2 across all test sets and number of interactions.

On the in-domain Test Set 1, MOIS-SAM2 achieved a scan-wise DSC of 0.60 with only three prompted lesions (single prompt per lesion).  In contrast, SAM2 started at a low DSC of 0.21 and improved slowly, plateauing below 0.33 even after 10 interactions. On Test Set 2, MOIS-SAM2 maintained a scan-wise DSC of approximately 0.50, while SAM2 exhibited slow gains, never exceeding 0.30. On the low tumor burden Test Set 3, SAM2 initially lagged behind but showed a steep rise in performance between 6 and 10 clicks, reaching a scan-wise DSC of 0.56. However, MOIS-SAM2 was already effective with fewer interactions and maintained a stable scan-wise DSC above 0.63 throughout. Finally, on Test Set 4, MOIS-SAM2 again outperformed SAM2 with fewer interactions, achieving 0.52 scan-wise DSC at saturation, compared to SAM2 (0.36 scan-wise DSC). These results illustrate the higher interaction efficiency of MOIS-SAM2, which consistently achieves stronger segmentation accuracy with fewer prompted lesions under various domain shift scenarios.

\subsection{Clinical Validity}
\label{subsec:clinical_validity_results}

\begin{figure}[t]
     \centering
     \includegraphics[width=0.6\linewidth]{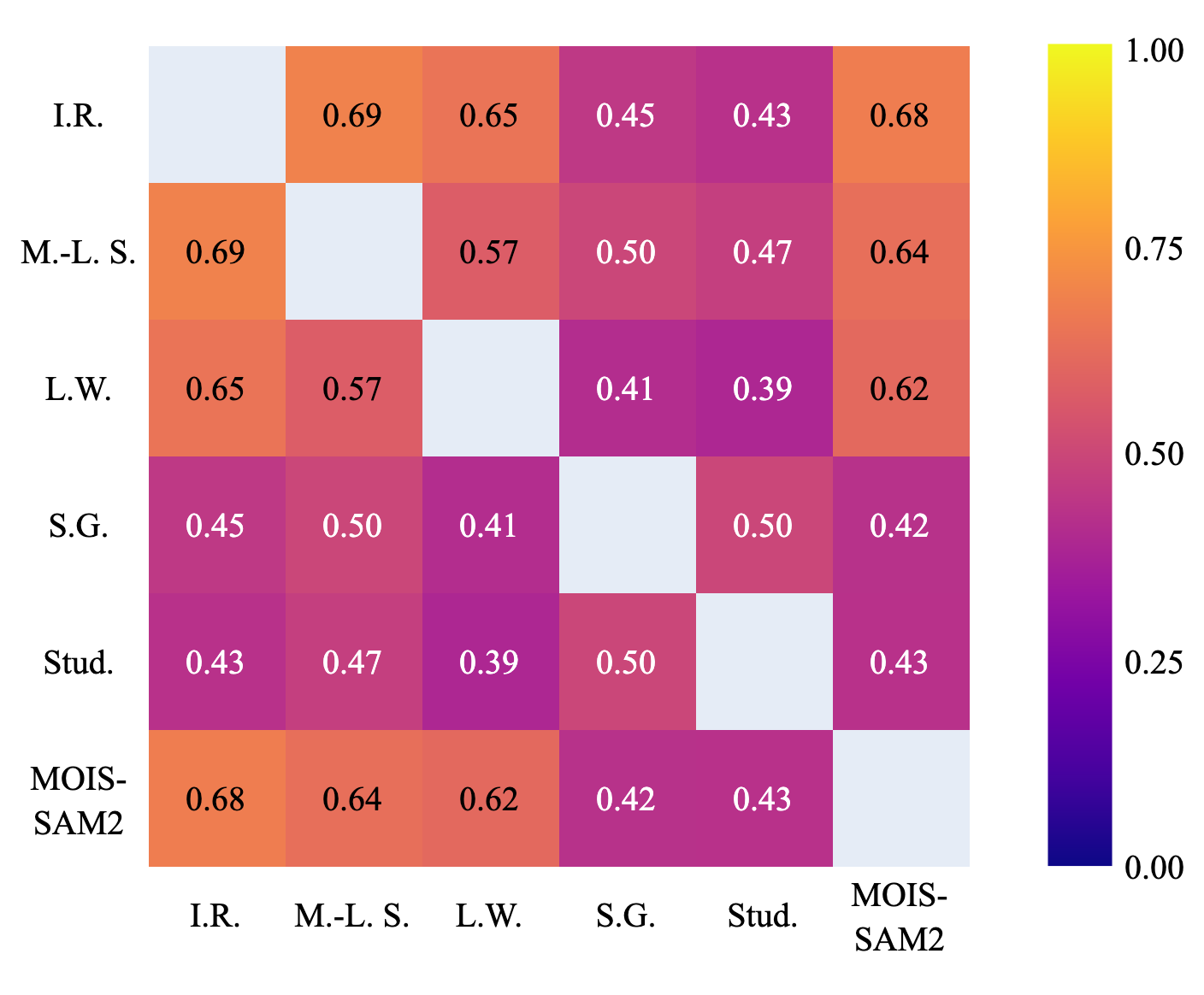}
     \caption{Scan-wise Dice Similarity Coefficient (DSC) heat map comparing segmentation results from five human annotators with varying radiological expertise and the proposed MOIS-SAM2 model. Each cell reflects the scan-wise DSC between two annotators or between an annotator and MOIS-SAM2, averaged across four representative NF1 patient cases. Annotators include three NF-specialized radiologists (I.R., M.-L.S., L.W.), one fourth year resident in radiology (S.G.), one medical student (Stud.), and the proposed MOIS-SAM2 model, supplied with 20 user clicks from a non-medical user.}
     \label{fig:Figure_6}
 \end{figure}

The inter-reader variability study for NF segmentation in T2w WB-MRI is presented in Figure \ref{fig:Figure_6}. The inter-reader agreement among expert radiologists (I.R., M.-L.S., L.W.) ranged from 0.57 to 0.69 in terms of scan-wise DSC, reflecting moderate consistency that highlights the inherent complexity of the NF segmentation task. Agreement between experts and less experienced annotators (S.G., medical student) dropped notably (e.g., I.R. vs. Stud.: 0.43), highlighting the challenge for non-experts to achieve clinically reliable segmentations. MOIS-SAM2 achieved scan-wise DSC scores of 0.62–0.68 when compared to expert annotations, placing it well within the inter-expert variability range. Importantly, this performance was achieved even though the model was prompted by a non-medical user using 20 clicks. This demonstrates that the model can approximate expert-level annotation quality. Its agreement with experts was consistently higher than that of both the fourth year resident in radiology and the medical student, further underscoring its clinical potential.

\section{Discussion}
\label{sec:discussion}

\subsection{Key Findings}
\label{subsec:key_findings}
This study aimed to address the central limitation in interactive segmentation of NFs in T2w WB-MRI of NF1 patients: the inability of existing models to scale to the dense, multi-lesion nature of the task without overwhelming the user with excessive interactions. We hypothesized that introducing exemplar-based semantic propagation into a transformer-based interactive segmentation framework SAM2 would enable accurate and efficient multi-lesion segmentation, substantially reducing user efforts while preserving lesion-wise accuracy.

The conducted experiments confirmed our hypotheses. The ablation study demonstrated that neither the unmodified SAM2 nor a naive reuse of the memory attention module could support exemplar-based semantic propagation. Instead, a dedicated exemplar attention mechanism with a structured exemplar representation was essential for achieving robust multi-lesion segmentation. This is because memory attention in SAM2 is designed to propagate a known object across adjacent slices, relying on strong spatial continuity. In contrast, exemplar-based semantic propagation requires matching semantically similar but spatially distant lesions across the scan. By introducing a dedicated exemplar attention module, trained to align features of semantically similar but spatially distant lesions, MOIS-SAM2 achieved robust multi-lesion segmentation.

We further showed that MOIS-SAM2 achieved strong segmentation performance even with a small number of prompted exemplars. As few as two exemplars in the semantic propagation significantly boosted accuracy, with performance saturating after 5–6 exemplars – highlighting the efficiency of semantic propagation. This reflects the ability of the model to generalize from a few user-segmented lesions to other semantically similar lesions across the scan. By leveraging shared visual-semantic representations, the model avoids the need to prompt each lesion individually, reducing interaction cost.

Compared to six baseline models – including automated convolution-based (U-Net, nnU-Net), interactive convolution-based (DINs, SW-FastEdit), and interactive transformer-based models (SAM2, VISTA3D) – MOIS-SAM2 achieved the highest scan-wise DSC and lesion detection F1 scores across all four test sets. Its superiority held even under domain shifts in MRI field strength, scanner vendor, and tumor burden. This robustness stems from the exemplar-based mechanism, which allows the model to adapt to the target domain by extracting and matching exemplars from the test scan itself.

In terms of interaction efficiency, MOIS-SAM2 consistently required fewer clicks to achieve higher segmentation performance than SAM2. The model also showed clinical validity: when guided by a non-clinical user with 20 clicks, MOIS-SAM2 matched the agreement levels seen among expert radiologists, and outperformed less experienced annotators.

Together, these results demonstrate that exemplar-based semantic propagation enables MOIS-SAM2 to effectively address the task of the scalable multi-lesion NF segmentation.

\subsection{Study Limitations}
\label{subsec:study_limitations}

Despite its strong performance, MOIS-SAM2 had several limitations.

The model introduced a separate exemplar attention module in addition to memory attention. While necessary for semantic propagation, this design increased architectural redundancy. Moreover, the exemplar and memory banks stored overlapping information, suggesting potential for a more efficient shared memory-exemplar bank. Currently, exemplar-based semantic propagation yielded a semantic mask with no lesion-level instance separation. Incorporating instance-aware outputs could enable downstream tracking or lesion-wise refinement. 

The ground truth was derived from a single annotator per scan. While inter-reader agreement among experts was in range 0.57 - 0.69 of scan-wise DSC, the absence of consensus masks limits robustness.

Although we performed a comprehensive evaluation on four diverse test sets, the study could benefit from broader validation on multi-center datasets with greater demographic and scanner diversity to confirm generalization.

\subsection{Implications}
\label{subsec:implications}

Our results demonstrate that exemplar attention enables generalization from a small set of user-prompted lesions to the full tumor burden. This exemplar-based logic can be integrated into the original SAM2 framework to accelerate segmentation of multiple same-class objects, reducing the need for repeated user prompting.

From a clinical perspective, current segmentation workflows for NF1 remain predominantly manual or rely on basic semi-automated techniques such as thresholding or region growing. While deep learning-based approaches like DINs offer improved segmentation accuracy \cite{zhang_dins_2022}, they are not integrated into widely used imaging platforms such as 3D Slicer \cite{fedorov_3d_2012}, limiting their accessibility in practice. Our model addresses this gap by being fully deployable within the 3D Slicer \cite{fedorov_3d_2012} via MONAI Label \cite{diaz-pinto_monai_2024}, facilitating direct testing and integration into radiological workflows. 

A potential barrier to the deployment of the proposed model in resource-constrained clinical environments is the requirement for GPU hardware during inference. Furthermore, the model usability, interaction efficiency, and clinical value must be validated in prospective studies involving end-users such as radiologists.

\subsection{Future Directions}
\label{subsec:future_directions}

Future work may focus on three main areas: architectural refinement, evaluation improvements, and clinical validation.

First, enabling instance-aware segmentation during exemplar-based propagation could support lesion tracking and refinement. Incorporating anatomical context or organ-aware prompting may help suppress false positives in commonly misclassified structures, such as lymph nodes or salivary glands, that share visual characteristics with NFs.

Second, the creation of consensus-based ground truth datasets and the adoption of standardized evaluation protocols – including consistent lesion size thresholds and metrics – would facilitate more robust training and cross-study comparisons.

Finally, prospective usability studies are needed to assess how radiologists could benefit from using MOIS-SAM2 in routine workflows. These studies should evaluate interaction time, correction frequency, trust in model outputs, inter-, and intra-reader variability.

\section{Conclusion}
We introduced MOIS-SAM2 for efficient multi-object interactive segmentation of NFs in T2w WB-MRI. By integrating exemplar-based semantic propagation into the transformer-base SAM2 framework, our model significantly reduced user efforts while maintaining high lesion-wise segmentation accuracy. MOIS-SAM2 outperformed existing automated and interactive baselines across diverse test sets. Deployed via MONAI Label \cite{diaz-pinto_monai_2024} in 3D Slicer \cite{fedorov_3d_2012}, the proposed method offers an accessible tool for interactive lesion segmentation in NF1 patients.

\section*{Ethics Statement}
This retrospective study was approved by the local ethics committee (2022-300201-WF approved on 19.04.2022, 2022-300201\_1-WF approved on 25.04.2022, 2022-300201\_2-WF approved on 23.06.2025) with a waiver of informed consent, adhering to data protection regulations and the Declaration of Helsinki.

\section*{CRediT authorship contribution statement}

\textbf{Georgii Kolokolnikov:} Conceptualization, Formal analysis, Investigation, Methodology, Software, Validation, Visualization, Writing - original draft. 

\textbf{Marie-Lena Schmalhofer:} Conceptualization, Data curation, Investigation, Validation, Writing - review \& editing.

\textbf{Sophie Götz:} Data curation, Investigation, Validation, Writing - review \& editing.

\textbf{Lennart Well:} Conceptualization, Data curation, Investigation, Validation, Writing - review \& editing.

\textbf{Said Farschtschi:} Data curation, Investigation, Resources, Validation, Writing - review \& editing.

\textbf{Victor-Felix Mautner:} Data curation, Investigation, Resources, Validation, Writing - review \& editing.

\textbf{Inka Ristow:} Conceptualization, Data curation, Funding acquisition, Investigation, Project administration, Validation, Writing - review \& editing.

\textbf{René Werner:} Conceptualization, Funding acquisition, Investigation, Methodology, Project administration, Supervision, Writing - review \& editing.

\section*{Declaration of Competing Interest}
The authors declare that they have no competing interests.

\section*{Acknowledgments}
This work was supported by the Deutsche Forschungsgemeinschaft (DFG SPP 2177, project number 515277218) and the German lay organization “Bundesverband Neurofibromatose e.V.” (grant/project no. n/a). The funders had no role in study design, data collection, model development, analysis, decision to publish, or preparation of the manuscript. The authors had full independence in all aspects of the study.

\section*{Supplementary Material}
The supplementary material includes a video demonstration (MP4, H.264 codec, 1080p resolution) of the proposed MOIS-SAM2 segmentation pipeline, showcasing exemplar-based interactive segmentation on two representative T2-weighted whole-body MRI scans of NF1 patients.

\bibliographystyle{elsarticle-num} 
\bibliography{Interactive_segmentation_bibliography}

\begin{thebibliography}{10}
\expandafter\ifx\csname url\endcsname\relax
  \def\url#1{\texttt{#1}}\fi
\expandafter\ifx\csname urlprefix\endcsname\relax\def\urlprefix{URL }\fi
\expandafter\ifx\csname href\endcsname\relax
  \def\href#1#2{#2} \def\path#1{#1}\fi

\bibitem{lammert_prevalence_2005}
M.~Lammert, et~al., Prevalence of {Neurofibromatosis} 1 in {German} {Children}
  at {Elementary} {School} {Enrollment}, Archives of Dermatology 141~(1)
  (2005).
\newblock \href {https://doi.org/10.1001/archderm.141.1.71}
  {\path{doi:10.1001/archderm.141.1.71}}.

\bibitem{thakur_multiparametric_2024}
U.~Thakur, et~al., Multiparametric whole-body {MRI} of patients with
  neurofibromatosis type {I}: spectrum of imaging findings, Skeletal Radiology
  (2024).
\newblock \href {https://doi.org/10.1007/s00256-024-04765-6}
  {\path{doi:10.1007/s00256-024-04765-6}}.

\bibitem{friedman_type_1997}
J.~M. Friedman, P.~H. Birch, Type 1 neurofibromatosis: {A} descriptive analysis
  of the disorder in 1728 patients, American Journal of Medical Genetics 70~(2)
  (1997) 138--143.
\newblock \href
  {https://doi.org/10.1002/(SICI)1096-8628(19970516)70:2<138::AID-AJMG7>3.0.CO;2-U}
  {\path{doi:10.1002/(SICI)1096-8628(19970516)70:2<138::AID-AJMG7>3.0.CO;2-U}}.

\bibitem{evans_malignant_2002}
D.~G.~R. Evans, Malignant peripheral nerve sheath tumours in neurofibromatosis
  1, Journal of Medical Genetics 39~(5) (2002) 311--314.
\newblock \href {https://doi.org/10.1136/jmg.39.5.311}
  {\path{doi:10.1136/jmg.39.5.311}}.

\bibitem{ahlawat_current_2020}
S.~Ahlawat, et~al., Current status and recommendations for imaging in
  neurofibromatosis type 1, neurofibromatosis type 2, and schwannomatosis,
  Skeletal Radiology 49~(2) (2020) 199--219.
\newblock \href {https://doi.org/10.1007/s00256-019-03290-1}
  {\path{doi:10.1007/s00256-019-03290-1}}.

\bibitem{heffler_tumor_2017}
M.~A. Heffler, et~al., Tumor segmentation of whole-body magnetic resonance
  imaging in neurofibromatosis type 1 patients: tumor burden correlates,
  Skeletal Radiology 46~(1) (2017) 93--99.
\newblock \href {https://doi.org/10.1007/s00256-016-2522-4}
  {\path{doi:10.1007/s00256-016-2522-4}}.

\bibitem{weizman_interactive_2012}
L.~Weizman, et~al., Interactive segmentation of plexiform neurofibroma tissue:
  method and preliminary performance evaluation, Medical \& Biological
  Engineering \& Computing 50~(8) (2012) 877--884.
\newblock \href {https://doi.org/10.1007/s11517-012-0929-1}
  {\path{doi:10.1007/s11517-012-0929-1}}.

\bibitem{isensee_nnu-net_2021}
F.~Isensee, et~al., {nnU}-{Net}: a self-configuring method for deep
  learning-based biomedical image segmentation, Nature Methods 18~(2) (2021)
  203--211.
\newblock \href {https://doi.org/10.1038/s41592-020-01008-z}
  {\path{doi:10.1038/s41592-020-01008-z}}.

\bibitem{zhang_dins_2022}
J.-W. Zhang, et~al., {DINs}: {Deep} {Interactive} {Networks} for {Neurofibroma}
  {Segmentation} in {Neurofibromatosis} {Type} 1 on {Whole}-{Body} {MRI}, IEEE
  Journal of Biomedical and Health Informatics 26~(2) (2022) 786--797.
\newblock \href {https://doi.org/10.1109/JBHI.2021.3087735}
  {\path{doi:10.1109/JBHI.2021.3087735}}.

\bibitem{diaz-pinto_deepedit_2022}
A.~Diaz-Pinto, et~al., {DeepEdit}: {Deep} {Editable} {Learning} for
  {Interactive} {Segmentation} of {3D} {Medical} {Images}, in: Data
  {Augmentation}, {Labelling}, and {Imperfections}, Springer Nature
  Switzerland, Cham, 2022, pp. 11--21.
\newblock \href {https://doi.org/https://doi.org/10.1007/978-3-031-17027-0_2}
  {\path{doi:https://doi.org/10.1007/978-3-031-17027-0_2}}.

\bibitem{hadlich_sliding_2024}
M.~Hadlich, et~al., Sliding {Window} {Fastedit}: {A} {Framework} for {Lesion}
  {Annotation} in {Whole}-{Body} {Pet} {Images}, in: 2024 {IEEE}
  {International} {Symposium} on {Biomedical} {Imaging} ({ISBI}), IEEE, Athens,
  Greece, 2024, pp. 1--5.
\newblock \href {https://doi.org/10.1109/ISBI56570.2024.10635459}
  {\path{doi:10.1109/ISBI56570.2024.10635459}}.

\bibitem{isensee_nninteractive_2025}
F.~Isensee, et~al., {nnInteractive}: {Redefining} {3D} {Promptable}
  {Segmentation}, arXiv:2503.08373 [cs.CV] (2025).
\newblock \href {https://doi.org/10.48550/ARXIV.2503.08373}
  {\path{doi:10.48550/ARXIV.2503.08373}}.

\bibitem{liu_simpleclick_2023}
Q.~Liu, et~al., {SimpleClick}: {Interactive} {Image} {Segmentation} with
  {Simple} {Vision} {Transformers}, in: 2023 {IEEE}/{CVF} {International}
  {Conference} on {Computer} {Vision} ({ICCV}), IEEE, Paris, France, 2023, pp.
  22233--22243.
\newblock \href {https://doi.org/10.1109/ICCV51070.2023.02037}
  {\path{doi:10.1109/ICCV51070.2023.02037}}.

\bibitem{kirillov_segment_2023}
A.~Kirillov, et~al., Segment {Anything}, in: 2023 {IEEE}/{CVF} {International}
  {Conference} on {Computer} {Vision} ({ICCV}), IEEE, Paris, France, 2023, pp.
  3992--4003.
\newblock \href {https://doi.org/10.1109/ICCV51070.2023.00371}
  {\path{doi:10.1109/ICCV51070.2023.00371}}.

\bibitem{ravi_sam_2024}
N.~Ravi, et~al., {SAM} 2: {Segment} {Anything} in {Images} and {Videos},
  arXiv:2408.00714 [cs] (2024).
\newblock \href {https://doi.org/10.48550/arXiv.2408.00714}
  {\path{doi:10.48550/arXiv.2408.00714}}.

\bibitem{he_vista3d_2024}
Y.~He, et~al., {VISTA3D}: {A} {Unified} {Segmentation} {Foundation} {Model}
  {For} {3D} {Medical} {Imaging}, arXiv:2406.05285 [cs.CV] (2024).
\newblock \href {https://doi.org/10.48550/ARXIV.2406.05285}
  {\path{doi:10.48550/ARXIV.2406.05285}}.

\bibitem{fedorov_3d_2012}
A.~Fedorov, et~al., {3D} {Slicer} as an {Image} {Computing} {Platform} for the
  {Quantitative} {Imaging} {Network}, Magnetic resonance imaging 30~(9) (2012)
  1323--1341.
\newblock \href {https://doi.org/10.1016/j.mri.2012.05.001}
  {\path{doi:10.1016/j.mri.2012.05.001}}.

\bibitem{diaz-pinto_monai_2024}
A.~Diaz-Pinto, et~al., {MONAI} {Label}: {A} framework for {AI}-assisted
  interactive labeling of {3D} medical images, Medical Image Analysis 95 (2024)
  103207.
\newblock \href {https://doi.org/10.1016/j.media.2024.103207}
  {\path{doi:10.1016/j.media.2024.103207}}.

\bibitem{solomon_automated_2004}
J.~Solomon, et~al., Automated detection and volume measurement of plexiform
  neurofibromas in neurofibromatosis 1 using magnetic resonance imaging,
  Computerized Medical Imaging and Graphics 28~(5) (2004) 257--265.
\newblock \href {https://doi.org/10.1016/j.compmedimag.2004.03.002}
  {\path{doi:10.1016/j.compmedimag.2004.03.002}}.

\bibitem{cai_tumor_2009}
W.~Cai, et~al., Tumor {Burden} in {Patients} with {Neurofibromatosis} {Types} 1
  and 2 and {Schwannomatosis}: {Determination} on {Whole}-{Body} {MR} {Images},
  Radiology 250~(3) (2009) 665--673.
\newblock \href {https://doi.org/10.1148/radiol.2503080700}
  {\path{doi:10.1148/radiol.2503080700}}.

\bibitem{weizman_pnist_2014}
L.~Weizman, et~al., {PNist}: interactive volumetric measurements of plexiform
  neurofibromas in {MRI} scans, International Journal of Computer Assisted
  Radiology and Surgery 9~(4) (2014) 683--693.
\newblock \href {https://doi.org/10.1007/s11548-013-0961-0}
  {\path{doi:10.1007/s11548-013-0961-0}}.

\bibitem{ho_image_2020}
C.~Y. Ho, et~al., Image segmentation of plexiform neurofibromas from a deep
  neural network using multiple b-value diffusion data, Scientific Reports
  10~(1) (2020) 17857.
\newblock \href {https://doi.org/10.1038/s41598-020-74920-1}
  {\path{doi:10.1038/s41598-020-74920-1}}.

\bibitem{wu_deep_2020}
X.~Wu, et~al., Deep {Parametric} {Active} {Contour} {Model} for
  {Neurofibromatosis} {Segmentation}, Future Generation Computer Systems 112
  (2020) 58--66.
\newblock \href {https://doi.org/10.1016/j.future.2020.05.001}
  {\path{doi:10.1016/j.future.2020.05.001}}.

\bibitem{wu_dh-gac_2022}
X.~Wu, et~al., {DH}-{GAC}: deep hierarchical context fusion network with
  modified geodesic active contour for multiple neurofibromatosis segmentation,
  Neural Computing and Applications (2022).
\newblock \href {https://doi.org/10.1007/s00521-022-07945-4}
  {\path{doi:10.1007/s00521-022-07945-4}}.

\bibitem{kiaei_development_2024}
D.~S. Kiaei, et~al., Development of a semi-automatic segmentation technique
  based on mean magnetic resonance imaging intensity thresholding for
  volumetric quantification of plexiform neurofibromas, Heliyon 10~(1) (2024).
\newblock \href {https://doi.org/10.1016/j.heliyon.2023.e23445}
  {\path{doi:10.1016/j.heliyon.2023.e23445}}.

\bibitem{kolokolnikov2024enhancing}
G.~Kolokolnikov, M.-L. Schmalhofer, I.~Ristow, R.~Werner,
  \href{https://openreview.net/forum?id=rSiCuOSOct}{Enhancing neurofibroma
  segmentation in whole-body {MRI}: Leveraging an anatomy-informed approach}
  (2024).
\newline\urlprefix\url{https://openreview.net/forum?id=rSiCuOSOct}

\bibitem{wasserthal_totalsegmentator_2023}
J.~Wasserthal, et~al., {TotalSegmentator}: {Robust} {Segmentation} of 104
  {Anatomic} {Structures} in {CT} {Images}, Radiology: Artificial Intelligence
  5~(5) (2023) e230024.
\newblock \href {https://doi.org/10.1148/ryai.230024}
  {\path{doi:10.1148/ryai.230024}}.

\bibitem{wei_multicenter_2025}
C.-J. Wei, et~al., A multicenter study of neurofibromatosis type 1 utilizing
  deep learning for whole body tumor identification, npj Digital Medicine 8~(1)
  (2025) 56.
\newblock \href {https://doi.org/10.1038/s41746-025-01454-z}
  {\path{doi:10.1038/s41746-025-01454-z}}.

\bibitem{rother_grabcut_2004}
C.~Rother, V.~Kolmogorov, A.~Blake, "{GrabCut}": interactive foreground
  extraction using iterated graph cuts, ACM Transactions on Graphics 23~(3)
  (2004) 309--314.
\newblock \href {https://doi.org/10.1145/1015706.1015720}
  {\path{doi:10.1145/1015706.1015720}}.

\bibitem{grady_random_2006}
L.~Grady, Random {Walks} for {Image} {Segmentation}, IEEE Transactions on
  Pattern Analysis and Machine Intelligence 28~(11) (2006) 1768--1783.
\newblock \href {https://doi.org/10.1109/TPAMI.2006.233}
  {\path{doi:10.1109/TPAMI.2006.233}}.

\bibitem{boykov_interactive_2001}
Y.~Boykov, M.-P. Jolly, Interactive graph cuts for optimal boundary \& region
  segmentation of objects in {N}-{D} images, in: Proceedings {Eighth} {IEEE}
  {International} {Conference} on {Computer} {Vision}. {ICCV} 2001, Vol.~1,
  IEEE Comput. Soc, Vancouver, BC, Canada, 2001, pp. 105--112.
\newblock \href {https://doi.org/10.1109/ICCV.2001.937505}
  {\path{doi:10.1109/ICCV.2001.937505}}.

\bibitem{xu_deep_2016}
N.~Xu, et~al., Deep {Interactive} {Object} {Selection}, in: 2016 {IEEE}
  {Conference} on {Computer} {Vision} and {Pattern} {Recognition} ({CVPR}),
  IEEE, Las Vegas, NV, USA, 2016, pp. 373--381.
\newblock \href {https://doi.org/10.1109/CVPR.2016.47}
  {\path{doi:10.1109/CVPR.2016.47}}.

\bibitem{maninis_deep_2018}
K.-K. Maninis, S.~Caelles, J.~Pont-Tuset, L.~Van~Gool, Deep {Extreme} {Cut}:
  {From} {Extreme} {Points} to {Object} {Segmentation}, in: 2018 {IEEE}/{CVF}
  {Conference} on {Computer} {Vision} and {Pattern} {Recognition}, IEEE, Salt
  Lake City, UT, USA, 2018, pp. 616--625.
\newblock \href {https://doi.org/10.1109/CVPR.2018.00071}
  {\path{doi:10.1109/CVPR.2018.00071}}.

\bibitem{faizov_interactive_2022}
B.~Faizov, V.~Shakhuro, A.~Konushin, Interactive {Image} {Segmentation} with
  {Transformers}, in: 2022 {IEEE} {International} {Conference} on {Image}
  {Processing} ({ICIP}), IEEE, Bordeaux, France, 2022, pp. 1171--1175.
\newblock \href {https://doi.org/10.1109/ICIP46576.2022.9897542}
  {\path{doi:10.1109/ICIP46576.2022.9897542}}.

\bibitem{wang_isegformer_2022}
Q.~Liu, Z.~Xu, Y.~Jiao, M.~Niethammer, {iSegFormer}: {Interactive}
  {Segmentation} via {Transformers} with {Application} to {3D} {Knee} {MR}
  {Images}, in: Medical {Image} {Computing} and {Computer} {Assisted}
  {Intervention} – {MICCAI} 2022, Vol. 13435, Springer Nature Switzerland,
  Cham, 2022, pp. 464--474.
\newblock \href {https://doi.org/10.1007/978-3-031-16443-9_45}
  {\path{doi:10.1007/978-3-031-16443-9_45}}.

\bibitem{wang2023sam}
H.~Wang, et~al., {SAM-Med3D}: Towards general-purpose segmentation models for
  volumetric medical images, in: Computer Vision -- ECCV 2024 Workshops,
  Springer Nature Switzerland, 2025, pp. 51--67.
\newblock \href {https://doi.org/https://doi.org/10.1007/978-3-031-91721-9_4}
  {\path{doi:https://doi.org/10.1007/978-3-031-91721-9_4}}.

\bibitem{zhu2024medical}
J.~Zhu, et~al., Medical sam 2: Segment medical images as video via segment
  anything model 2, arXiv preprint arXiv:2408.00874 (2024).
\newblock \href {https://doi.org/https://doi.org/10.48550/arXiv.2408.00874}
  {\path{doi:https://doi.org/10.48550/arXiv.2408.00874}}.

\bibitem{chen_transunet_2024}
J.~Chen, et~al., {TransUNet}: {Rethinking} the {U}-{Net} architecture design
  for medical image segmentation through the lens of transformers, Medical
  Image Analysis 97 (2024) 103280.
\newblock \href {https://doi.org/10.1016/j.media.2024.103280}
  {\path{doi:10.1016/j.media.2024.103280}}.

\bibitem{rana_dynamite_2023}
A.~K. Rana, S.~Mahadevan, A.~Hermans, B.~Leibe, {DynaMITe}: {Dynamic} {Query}
  {Bootstrapping} for {Multi}-object {Interactive} {Segmentation}
  {Transformer}, in: 2023 {IEEE}/{CVF} {International} {Conference} on
  {Computer} {Vision} ({ICCV}), IEEE, Paris, France, 2023, pp. 1043--1052.
\newblock \href {https://doi.org/10.1109/ICCV51070.2023.00102}
  {\path{doi:10.1109/ICCV51070.2023.00102}}.

\bibitem{li2024learning}
K.~Li, H.~Cheng, G.~Vosselman, M.~Y. Yang, Learning from exemplars for
  interactive image segmentation, arXiv preprint arXiv:2406.11472 (2024).
\newblock \href {https://doi.org/https://doi.org/10.48550/arXiv.2406.11472}
  {\path{doi:https://doi.org/10.48550/arXiv.2406.11472}}.

\bibitem{wu_one-prompt_2024}
J.~Wu, M.~Xu, One-{Prompt} to {Segment} {All} {Medical} {Images}, in: 2024
  {IEEE}/{CVF} {Conference} on {Computer} {Vision} and {Pattern} {Recognition}
  ({CVPR}), IEEE, Seattle, WA, USA, 2024, pp. 11302--11312.
\newblock \href {https://doi.org/10.1109/CVPR52733.2024.01074}
  {\path{doi:10.1109/CVPR52733.2024.01074}}.

\bibitem{tejani_checklist_2024}
A.~S. Tejani, et~al., Checklist for {Artificial} {Intelligence} in {Medical}
  {Imaging} ({CLAIM}): 2024 {Update}, Radiology: Artificial Intelligence 6~(4)
  (2024) e240300.
\newblock \href {https://doi.org/10.1148/ryai.240300}
  {\path{doi:10.1148/ryai.240300}}.

\bibitem{noauthor_neurofibromatosis_1988}
{National Institutes of Health}, Neurofibromatosis. {Conference} statement.
  {National} {Institutes} of {Health} {Consensus} {Development} {Conference},
  Archives of Neurology 45~(5) (1988) 575--578.

\bibitem{marinov_deep_2024}
Z.~Marinov, et~al., Deep {Interactive} {Segmentation} of {Medical} {Images}:
  {A} {Systematic} {Review} and {Taxonomy}, IEEE Transactions on Pattern
  Analysis and Machine Intelligence 46~(12) (2024) 10998--11018.
\newblock \href {https://doi.org/10.1109/TPAMI.2024.3452629}
  {\path{doi:10.1109/TPAMI.2024.3452629}}.

\end{thebibliography}
\end{document}